\documentclass[twocolumn,showpacs,pra,aps,superscriptaddress]{revtex4-1}
\usepackage{array}
\usepackage{amsmath}
\usepackage{graphicx}
\usepackage{epstopdf}
\usepackage{color}
\usepackage{amsfonts}
\usepackage{subfigure}

\def\bra#1{\mathinner{\langle{#1}|}} 
\def\ket#1{\mathinner{|{#1}\rangle}}

\providecommand{\abs}[1]{\lvert#1\rvert}  

\def\e{\mathrm{e}}
\def\ii{\mathrm{i}}

\newcommand{\onecolm}{
  \end{multicols}
  \vspace{-3.5ex}
  \noindent\rule{0.5\textwidth}{0.1ex}\rule{0.1ex}{2ex}\hfill
}
\newcommand{\twocolm}{
  \hfill\raisebox{-1.9ex}{\rule{0.1ex}{2ex}}\rule{0.5\textwidth}{0.1ex}
  \vspace{-4ex}
  \begin{multicols}{2}
}

\begin{document}
\title{Statistical mechanics of the Cluster-Ising model}
\date{\today}

\author{Pietro Smacchia}
\affiliation{SISSA - via Bonomea 265, 34136, Trieste, Italy}
\author{Luigi Amico}
\affiliation{ CNR-MATIS-IMM $\&$ Dipartimento di Fisica e  Astronomia Universit\`{a} di Catania, C/O ed. 10, viale A. Doria 6 95125 Catania, Italy}
\author{Paolo Facchi}
\affiliation{Dipartimento di Matematica and MECENAS, Universit\`a di Bari, I-70125 Bari, Italy}
\affiliation{INFN, Sezione di Bari, I-70126 Bari, Italy}

\author{Rosario Fazio}
\affiliation{NEST, Scuola Normale Superiore and Istituto Nanoscienze -- CNR,  56126 Pisa, Italy}
\affiliation{Center for Quantum Technology, National University of Singapore,
	117542 Singapore, Singapore}

\author{Giuseppe Florio}
\affiliation{Dipartimento di Fisica and MECENAS, Universit\`a di Bari, I-70126 Bari, Italy}
\affiliation{INFN, Sezione di Bari, I-70126 Bari, Italy}

\author{Saverio Pascazio}
\affiliation{Dipartimento di Fisica and MECENAS, Universit\`a di Bari, I-70126 Bari, Italy}
\affiliation{INFN, Sezione di Bari, I-70126 Bari, Italy}

\author{Vlatko Vedral}
\affiliation{Center for Quantum Technology, National University of Singapore,
	117542 Singapore, Singapore}
\affiliation{Department of Physics, National University of
	Singapore, 2 Science Drive 3, Singapore 117542}
\affiliation{Department of Physics, University of Oxford, Clarendon Laboratory, Oxford, OX1 3PU, UK}

\begin{abstract}
We study a Hamiltonian system describing a three-spin-$1/2$ cluster-like interaction competing with an Ising-like anti-ferromagnetic interaction. We compute  free energy, spin correlation functions  and entanglement both in  the ground and in thermal states.  The model undergoes a quantum phase transition between an Ising phase with a nonvanishing magnetization and a cluster phase characterized by a string order. 
Any two-spin entanglement is found to vanish in both quantum phases because of a nontrivial correlation pattern. Neverthless, the residual multipartite entanglement is maximal in the cluster phase and dependent on the magnetization in the Ising phase. We  study the block entropy at the critical point and calculate the central charge of the system, showing that the criticality of the system is beyond the Ising universality class.
\end{abstract}

\maketitle

\section{Introduction}

The interplay between quantum information and statistical mechanics has given rise to a  new trend in contemporary physics research. On one hand, quantum informatics  provides new views into statistical physics, with spin-offs that could lead to decisive progress in the field; on the other hand,  strongly correlated quantum statistical systems, considered as platforms for future quantum computers, naturally provide correlations that  quantum information aims to exploit  as a resource \cite{ladd,freedman}. Cold-atom quantum simulators play an important role in this context because they provide quantum statistical systems, beyond traditional condensed matter realizations \cite{coldatoms}.  Within this interdisciplinary field, quantitative analysis of the entanglement encoded in a given state of the statistical system provides precious informations on its physical properties \cite{manybodyreview,calabresecardy,cramer_rev}.

This article deals with a one dimensional statistical system formulated in the cross-fertilization area described above. The reference system is provided by cold atoms in a triangular optical lattice \cite{triangular}. For a suitable choice of the parameters, such a system can be considered as a  spin system with a specific  ring-exchange interaction in a triangular lattice that, in turn,  can be recast into a ``zig-zag chain''.  Remarkably, the ground state of the system is the so-called cluster state \cite{Pachos04}.  The  setup provides a physical platform for  the one-way route to quantum computation, where the algorithm consists in suitable measurements aiming at  reconstructing the  high degree of entanglement  characterizing the cluster state \cite{Briegel01}. Interestingly, besides the three-spin ring-exchange interaction, various two-spin interactions can be shown to emerge in the system. Therefore, the cluster interaction competes with the exchange one by tuning a control parameter \cite{Pachos04}. The interplay between these two interactions has been independently considered in the context of quantum information, in order  to estimate the effects of local perturbations on the cluster states \cite{Doherty09,Skrovseth09,Li}.  It results  that the correlation pattern characterizing the cluster state  is robust up to a critical value of the control parameter, meaningfully defining a ``cluster phase"; for larger values of the control parameter the system is in an Ising phase. The two phases are separated by a continous quantum phase transition (QPT) \cite{sachdevbook}. Interestingly enough, the cluster phase is characterized by a diverging range of localizable entanglement \cite{cirac-localizable} that can be traced back to a non-vanishing so-called string order parameter \cite{camposvenuti}.

Similar findings were recently obtained for a slight variation of the model discussed above, that we shall call cluster-Ising model (CIM), whose Hamiltonian reads \cite{primoart}
\begin{equation}
H(\lambda)=-\sum_{j=1}^N \sigma_{j-1}^x \sigma_j^z \sigma_{j+1}^x + \lambda \sum_{j=1}^N \sigma_j^y \sigma_{j+1}^y,
\label{eq:ham_cluster}
\end{equation}
where $\sigma_i^{\alpha}$, $\alpha=x,y,z$, are the Pauli matrices and, except otherwise stated, we take periodic boundary conditions $\sigma^{\alpha}_{N+k}= \sigma^{\alpha}_{k}$. We note  that the nearest-neighbor and the next-to-nearest-neighbor exchange in the two- and three-spin interaction, respectively, involve orthogonal spin orientations (in contrast with the model studied in  \cite{Doherty09,Skrovseth09}); therefore all the spin components participate to the interaction.

In this article we intend to study the statistical mechanical features of this system, by analyzing correlation functions, entanglement and quantum phase transitions. We shall focus in particular on the remarkable interplay between the competing ``cluster" and ``Ising" phases.

\section{Overview}

\begin{table*}
\caption{Overview of the properties of the ground state as $\lambda$ varies}
\label{table2}
\begin{tabular}{ | c |c | c|}
\hline
Cluster state $@ \lambda=0$ & QPT $@ \lambda=1$ &  Ising state $@\lambda=\infty $\\
\hline  
   symmetry protected topological order & central charge $c=3/2$   & $Z_2 $ symmetry \\
  by $Z_2\times Z_2 $ symmetry &    $\quad  \nu$=z=1, $\beta=3/8, \alpha=0$   &  \\
\hline 
\end{tabular}
\end{table*}

For $\lambda=0$ the ground state is a cluster state $\ket{C}$, defined as  the unique common eigenvector of the set of commuting Hermitian operators (known as ``stabilizers'') $K_j=\sigma_{j-1}^x\sigma_j^z\sigma_{j+1}^x $: $K_j \ket{C} =+1  \ket{C}$ \cite{hein}.
The quench of the cluster to antiferromagnetic order occurs through an exotic quantum phase transition from an antiferromagnet to a phase with a hidden  order - the cluster state provides an example of Symmetry Protected Topological Order (SPTO)  by a $Z_2\times Z_2$ symmetry \cite{primoart,wen-protected,AKLT-protected} (see \cite{miyake} for applications of the notion of SPTO to quantum computation). 
In the present article we shall corroborate these findings by studying the degeneracy of the ground state of the system through a mapping of the spins onto Majorana fermions (see Fig.~\ref{fig:majorana}). For the cluster state $\lambda=0$ the degeneracy of the ground state is due to four uncoupled Majorana fermions.  

We shall obtain in Sec.~\ref{sec:corr_functions} the partition function of the system together with the exact expressions of the correlation function $R_{jl}^\gamma(T)=\langle \sigma_j^\gamma \sigma_l^\gamma\rangle_T$, with $\gamma=x,y,z$,  and $T$ temperature, in terms of Toeplitz determinants.  A nontrivial pattern of spin-spin correlation functions emerge: $R_{jl}^z(T)$, together with the magnetization along $z$ identically vanish; $R_{jl}^x(T)$ identically vanishes, unless $|j-l|$ is a multiple of $3$;  $R_{jl}^y(T)$ is nonvanishing. 

\begin{figure}
\begin{center}
\includegraphics[width=\columnwidth]{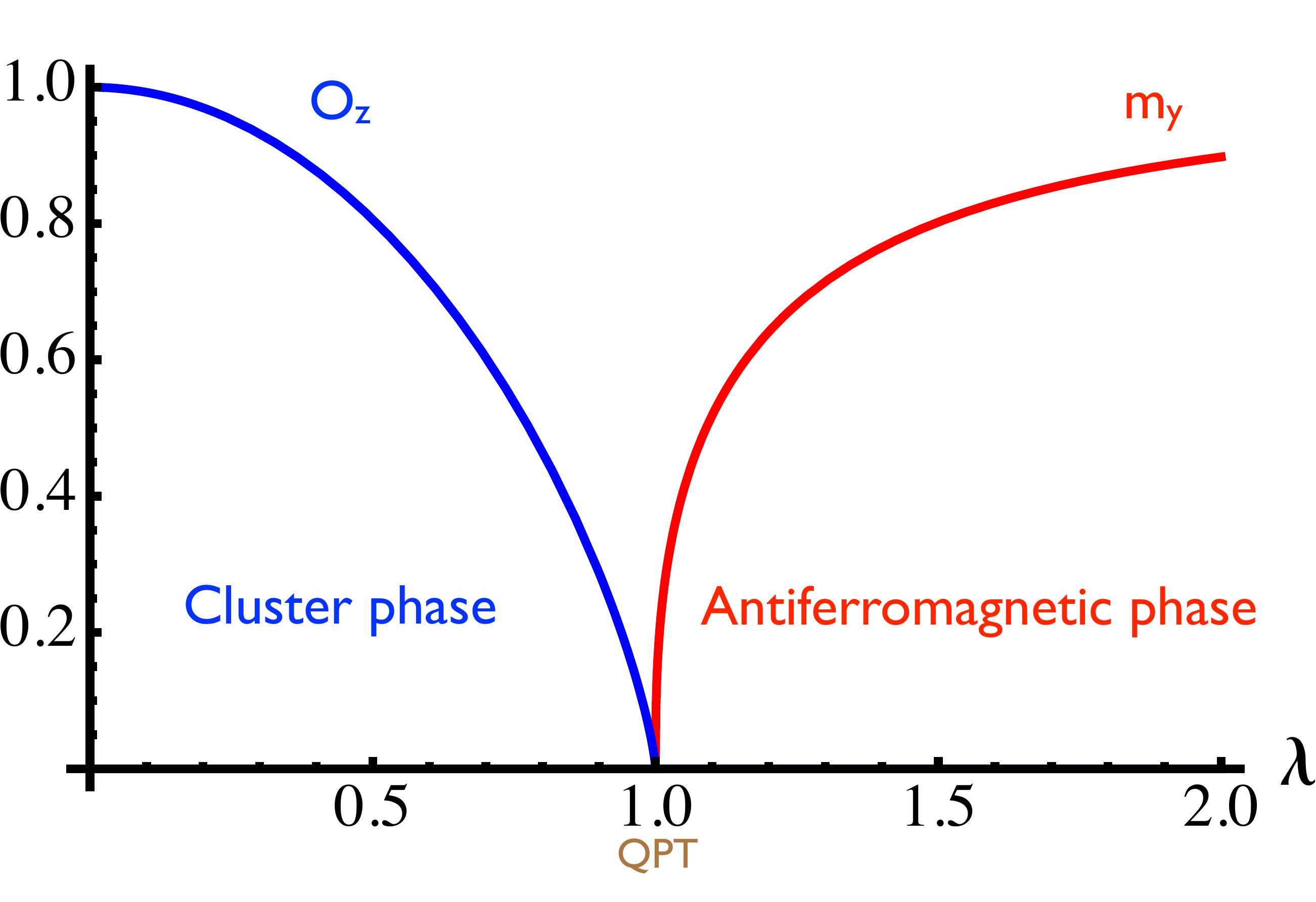}
\caption{(Color online) String order parameter $O_z$ (blue) and staggered magnetization $m_y$ (red) versus $\lambda$. Their behavior clearly shows the existence of two different phases of the system: when $\lambda<1$ there is topological order protected by the symmetry $Z_2\times Z_2 $ and a non-vanishing string order parameter, while when $\lambda>1$ there is antiferromagnetic order and a non-vanishing staggered magnetization.}
\label{fig:order_parameters}
\end{center}
\end{figure}

The cluster and the Ising phases are characterized by the string order parameter $O_z$ and the staggered magnetization $m_y$, respectively. See Fig.~\ref{fig:order_parameters}. Both will be obtained exactly in Secs.~\ref{sec:staggered_m} and \ref{sect:string}. 
In particular, we find $O_z\sim \left(1-\lambda\right)^{3/4}$ and $\displaystyle{m_y\sim(\lambda-1)^{\frac{3}{8}} }$, showing that the quantum phase transition is not in the Ising universality class (that would have yielded $\beta=1/8$).

\begin{figure}
\includegraphics[width=\columnwidth]{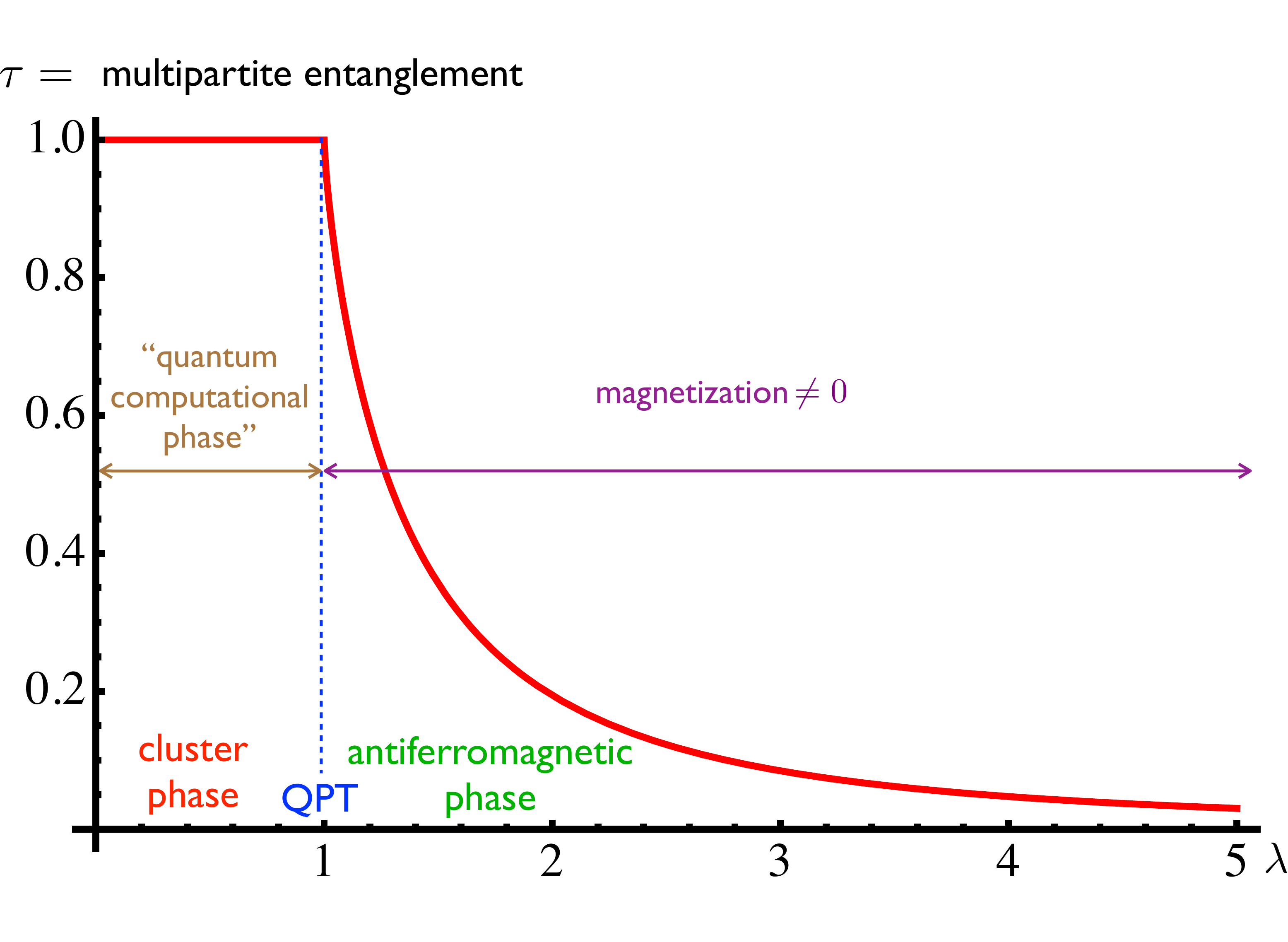}
\caption{(Color online) Residual (multipartite) entanglement $\tau$ in the symmetry breaking ground state versus $\lambda$. From its behavior we can identify the two phases of the system: a ``cluster phase'' (useful for quantum computation), where multipartite entanglement is constantly equal to its maximum value,  and an anti-ferromagnetic phase, characterized by a sharp decrease in $\tau$ and the presence of a non-vanishing staggered magnetization along the $y$-axis.}
\label{fig:ent_res}
\end{figure}

The entanglement pattern will be investigated in Sec.~\ref{sec:entanglement}. Two-spin entanglement identically vanishes, both in the ``thermal" ground state and in the ground state with broken symmetry, both at $T=0$ and at any finite temperature. It is therefore unable to detect the quantum phase transition. Nevertheless residual (multipartite) entanglement $\tau$ is nonvanishing and is viewed as a global figure of merit. We 
find  that it saturates at its maximal value  $\tau=1$ for any $\lambda$ in the ``thermal" ground state. 
However, interestingly, when one looks at the symmetry-breaking ground state, multipartite entanglement saturates in the whole cluster phase, but it decreases for $\lambda >1$, marking the critical point with $\partial_{\lambda} \tau(\lambda) \stackrel{\lambda \rightarrow 1^+}{\sim} -{\rm cost}\times(\lambda-1)^{-1/4}$. See Fig.~\ref{fig:ent_res}. The multipartite entanglement of the symmetry-breaking ground state is therefore able to detect the phase transition.

The block entropy is studied in Sec.~\ref{sec:entropy}
(see Fig.~\ref{fig:entropy}) and yields $c=3/2$ as central charge. We demonstrate that such a result can be traced back to the periodicity of the free energy, implying in turn that the system is energetically equivalent to three uncoupled Ising chains. Therefore, the criticality of CIM is characterized by an emergent 
$E_8\times E_8\times E_8$ symmetry \cite{E_8}.

Summarizing: for $\lambda=0$ and open boundary conditions, the system has a $Z_2\times Z_2$ symmetry that protects the  topological order \cite{primoart}; for $\lambda<1$ there is a ``cluster phase" with a non-local hidden order, detected by a nonvanishing string-order parameter $O_z$; for $\lambda>1$ the system is antiferromagnetic with long-range order. The antiferromagnetic and the cluster phases are separated by a continuous quantum phase transition.

The critical indices of the quantum phase transition are summarized in Table \ref{table2} (where $\nu$, z and $\alpha$ are related to the correlation length, the dynamical correlation functions and the specific heat, respectively). 
Finally, in Sec.~\ref{sec:conclusions} we compare our findings with other known models with three-spin interactions
and draw some conclusions.

\section{The exact solution}
\label{sec:model}

\subsection{Diagonalization}
Despite the presence of a three-spin interaction, the Hamiltonian (\ref{eq:ham_cluster}) can be diagonalized and describes free fermions. Introducing the Jordan-Wigner transformations 
 \begin{equation}
 \label{eq:jw}
 c_j =\left( \prod_{m=1}^{j-1} \sigma_m^z \right) \sigma^-_j, \quad c^\dagger_j =\left( \prod_{m=1}^{j-1} \sigma_m^z \right) \sigma^+_j,
 \end{equation}
where $\sigma^{\pm}_j= (\sigma^x_j \pm \ii \sigma^y_j)/2$,
Eq.~(\ref{eq:ham_cluster}) takes the form
\begin{eqnarray}
\label{eq:fermihamiltonian}
H(\lambda)&=& \sum_{l=1}^{N} (c_{l-1}^\dagger-c_{l-1})(c_{l+1}^\dagger+c_{l+1})
\nonumber \\ 
& & + \lambda \sum_{l=1}^{N} (c_{l}^\dagger+c_{l})(c_{l+1}^\dagger-c_{l+1}),
\end{eqnarray}
apart from a border term, which is negligible in the thermodynamic limit.
Furthermore, we apply a Fourier transformation
$
b_k=\frac{1}{\sqrt{N}} \sum_{j=1}^N \e^{-\frac{2 \pi \ii k j}{N}} c_j,
$ 
with $k=1,\dots,N$,
followed by a Bogoliubov transformation,
$ 
 b_k  = u_k \gamma_k + \ii v_k \gamma^{\dagger}_{-k},
$ 
with
\begin{equation}
u_k=\frac{1}{\sqrt{2}} \sqrt{1+\frac{\epsilon_k}{\Lambda_k}} , \quad
v_k=-\frac{1}{\sqrt{2}}\,{\rm sign} (\delta_k) \sqrt{1-\frac{\epsilon_k}{\Lambda_k}},
\end{equation}
\begin{equation} 
\left\{
\begin{aligned}
& \epsilon_k = \cos\left(\frac{4 \pi k}{N} \right)-\lambda \cos\left(\frac{2 \pi k}{N} \right), \\
&\delta_k=\sin \left(\frac{4 \pi k}{N} \right) + \lambda \sin\left(\frac{2 \pi k}{N} \right),
\end{aligned}
\right. 
\end{equation}
and
$ 
\Lambda_k=\sqrt{1+\lambda^2-2 \lambda \cos\left(\frac{6 \pi k}{N} \right)}. 
$ 
Finally, we obtain
\begin{equation}
H(\lambda)=2 \sum_{k=1}^N \Lambda_k \left( \gamma^\dagger_k \gamma_k-\frac{1}{2}\right),
\label{eq:diagonal_form}
\end{equation}
whose ground state is defined by
$\gamma_k\ket{\Omega}=0$, $\forall k$. 

\subsection{Free Energy}

From Eq.~(\ref{eq:diagonal_form}) we can compute the partition function of the system:
\begin{equation}
\begin{split}
\mathcal{Z}(\beta,\lambda)&= {\rm Tr}\, e^{-2 \beta \sum_k \Lambda_k \left(\gamma^{\dagger}_k \gamma_k-1/2\right)}  \\
& = \prod_k 2 \cosh \left(\beta \Lambda_k\right),
\end{split}
\end{equation}
where $\beta=1/k_BT$ with $k_B$ being the Boltzmann constant and $T$ the temperature of the system.  In the  thermodynamic limit ($N \rightarrow \infty$), the free energy density reads
\begin{eqnarray}
f(\beta,\lambda) = -\frac{1}{\pi \beta} \int_0^\pi dp \log\left[2 \cosh \left(\beta \Lambda(p)\right)\right],
\label{free_ising}
\end{eqnarray}
where
\begin{equation}\label{eq:dispersion}
\Lambda(p)=\sqrt{1+\lambda^2-2 \lambda \cos (3p)}.
\end{equation}

Notice now that, by using the periodicity of the dispersion relation $\Lambda(p)$, the free energy~(\ref{free_ising}) of the CIM~(\ref{eq:ham_cluster}) can be rewritten as
\begin{equation}
f(\beta,\lambda) = -\frac{1}{\pi \beta} \int_0^\pi dp \log\left[2 \cosh \left(\beta \Lambda^{\mathrm{Ising}}(p)\right)\right],
\label{free_ising1}
\end{equation}
where
$ 
\Lambda^{\mathrm{Ising}}(p)=\sqrt{1+\lambda^2-2 \lambda \cos p}
$ 
is the dispersion relation of 
the quantum  Ising chain in a transverse field:
\begin{equation}
H_{\rm Ising}(\lambda) = - \sum_{j=1}^{N} \left[\sigma_j^x \sigma_{j+1}^x+\lambda \sigma_j^z\right] .
\label{eq:ham_ising}
\end{equation}
Therefore, the system has exactly  \emph{the same} free energy of a suitable Ising model where the cluster interaction and the antiferromagnetic exchange along $y$ turn into an effective antiferromagnetic exchange along $x$ and an external field, respectively.

According to the general theory of continuous phase transitions, the second derivative of the free energy density with respect to $\lambda$ is divergent:
\begin{eqnarray}
\partial_{\lambda}^2 f(\beta,\lambda) & \stackrel{\beta \rightarrow \infty}{\rightarrow} &  \frac{1}{\pi \lambda^2 ( 1+\lambda)} \left[\left(\lambda+1\right)^2 \mathcal{E}\left(\frac{4 \lambda}{(1+\lambda)^2} \right) \right. \nonumber \\
& & \left. - \left(1+\lambda^2 \right) \mathcal{K} \left(\frac{4 \lambda}{(1+\lambda)^2} \right)\right] \nonumber \\
& \stackrel{\lambda \to 1}{\sim}& {\rm cost} \times\log \abs{\lambda-1},
\label{eq:derivata_seconda}
\end{eqnarray}
where $\mathcal{E}$ and $\mathcal{K}$ are the complete elliptic integral of the second and  first kind, respectively:
\begin{equation} 
\left\{
\begin{aligned}
&\mathcal{E}(x)=\int_0^{\frac{\pi}{2}} \sqrt{1-x^2 \sin^2 \theta} \, d \theta ,\\
&\mathcal{K}(x)=\int_0^{\frac{\pi}{2}} \frac{d \theta}{\sqrt{1-x^2 \sin^2 \theta}} ,
\end{aligned}
\right.
\end{equation}
and the divergence of Eq.~(\ref{eq:derivata_seconda}) in the limit $\lambda \rightarrow 1$  is a consequence of the singular behavior of $\mathcal{K}(x)$ at $ x=1$. Such a singular behavior is ultimately due to the vanishing of the energy gap  between the ground  and the first excited state  at the critical mode $p=0$ and $\lambda=1$, with critical indices z = $\nu=1$.

\subsection{Duality}
\label{sect:duality}

The above-discussed link with the Ising model  indicates that the system must enjoy a nontrivial  duality.
Indeed, consider the duality transformation
\begin{equation}
\label{eq:dual}
\mu_{j}^z =\sigma_j^x\sigma_{j+1}^x, \quad 
\mu_{j}^x =\prod_{k=1}^{j}\sigma_{k}^z
\end{equation}
(with the convention $\sigma_{N+1}^x=1$), where $\mu^\alpha_j$ are  the Pauli matrices.
Since the inverse of the second transformation is $\sigma_j^z = \mu_{j-1}^x \mu_j^x$ (with $\mu_0^x =1$),
one obviously gets from~(\ref{eq:ham_ising})
\begin{equation}
\label{eq:dualIs}
H^{\text{dual}}_{\rm Ising} (\lambda)=\lambda H_{\rm Ising}(\lambda^{-1}),
\end{equation}
apart from a boundary term. But one also gets that the CIM Hamiltonian (\ref{eq:ham_cluster}) behaves in the same way:
\begin{equation}
\label{eq:dual2}
H^{\text{dual}}(\lambda)=\lambda H(\lambda^{-1}).
\end{equation}
Thus, the Ising interaction is mapped into the cluster one and \emph{viceversa},
when 
the thermodynamic limit is considered. 
We notice that the QPT $\lambda=1$ corresponds to a self-dual point \cite{Doherty09,Skrovseth09,primoart}.

\subsection{Majorana fermions}
\label{sect:majorana}
\begin{figure}
\begin{center}
\includegraphics[width=0.9\columnwidth]{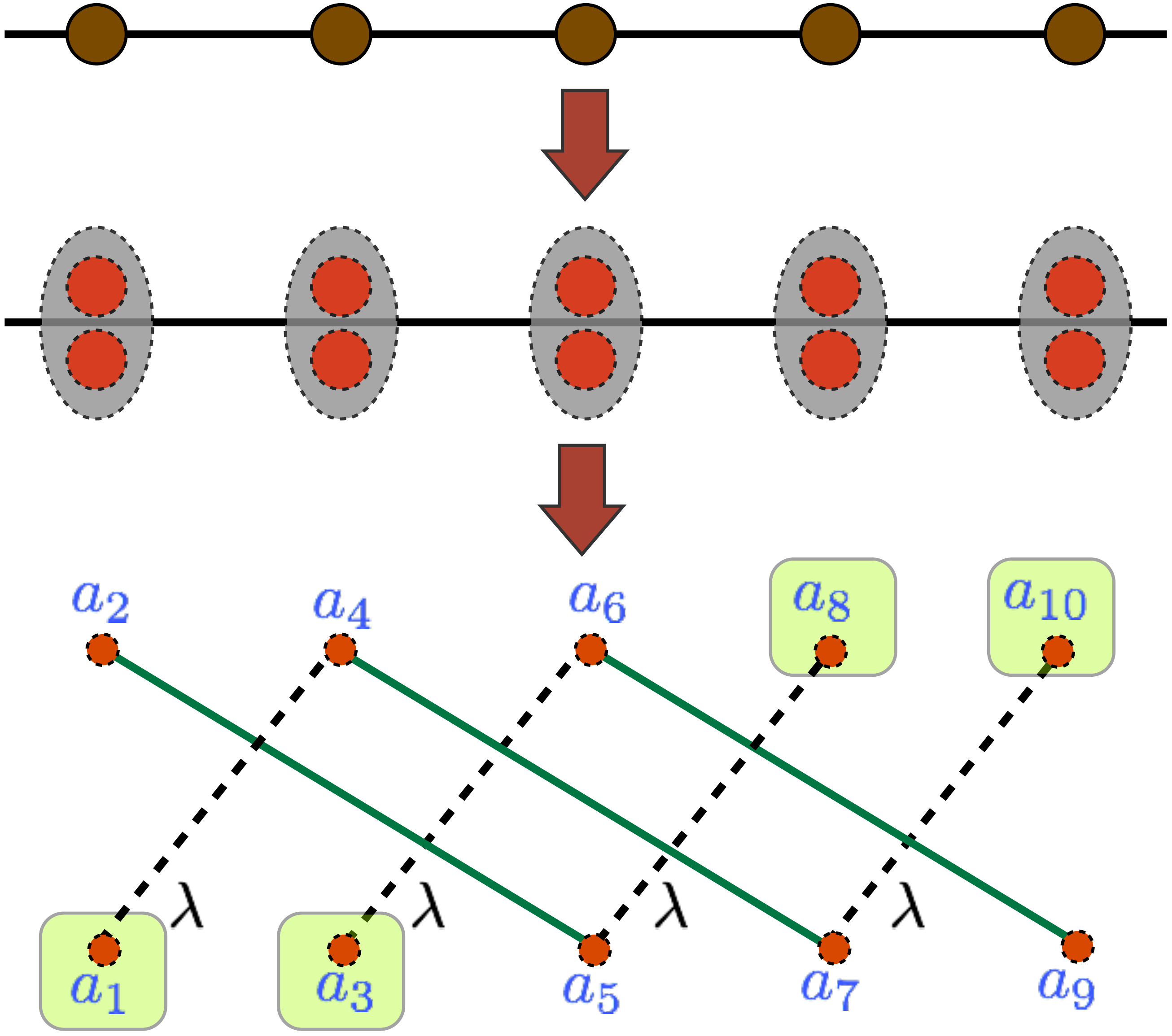}
\caption{(Color online) Each  fermion of the chain (top) can be represented by a Majorana pair
(center). The Majorana fermions can be arranged according to the
interaction described by the Hamiltonian (\ref{eq:ham-majorana}) (bottom). The (black) dashed lines represent Ising interactions. The (green) continuous lines represent the cluster interactions. The boxed Majorana fermions are responsible for the ground state degeneracy of the cluster state.}
\label{fig:majorana}
\end{center}
\end{figure}

An additional piece of information is unveiled  when one expresses the CIM Hamiltonian (\ref{eq:ham_cluster}) in terms of  Majorana fermions (see Fig.~\ref{fig:majorana}). In terms of the fermionic operators introduced by the Jordan-Wigner transformations (\ref{eq:jw}), we have
\begin{equation}
\label{majorana_fermions}
\check{a}_{2j-1}=c^\dagger_j+c_j, \quad \check{a}_{2j}= \ii (c_j-c^\dagger_j),
\end{equation}
($j=1,\dots,N$), with 
\begin{equation}
\check{a}_k^\dagger=\check{a}_k,\quad\left\{\check{a}_k,\check{a}_l\right\}= 2 \delta_{kl},
\end{equation}
($k,l=1,\dots,2N$).
The CIM Hamiltonian becomes 
\begin{equation}\label{eq:ham-majorana}
H(\lambda)=\ii \sum_{l=1}^{N} \check{a}_{2l-2}\check{a}_{2l+1} +\ii \lambda\sum_{l=1}^{N} \check{a}_{2l-1}\check{a}_{2l+2} ,
\end{equation}
where periodic boundary conditions have been considered. 
Notice that in  an open chain 
one gets
\begin{equation}\label{eq:ham-majorana1}
H(\lambda)=\ii \sum_{l=2}^{N-1} \check{a}_{2l-2}\check{a}_{2l+1} +\ii \lambda\sum_{l=1}^{N-1} \check{a}_{2l-1}\check{a}_{2l+2},
\end{equation}
so that, by turning off the Ising interaction, 
four free Majorana fermions emerge $\check{a}_{1}, \check{a}_{3}, \check{a}_{N-2}, \check{a}_{N}$  (Fig.~\ref{fig:majorana}).  This effect marks the non trivial  $Z_2\times Z_2$ of  the ground-state degeneracy of the system in the cluster limit $\lambda=0$.
by tuning off the Ising interaction in  an open chain 
four free Majorana fermions emerge $\check{a}_{1}, \check{a}_{3}, \check{a}_{N-2}, \check{a}_{N}$  (Fig.~\ref{fig:majorana}).  This effect marks the non trivial  $Z_2\times Z_2$ of  the ground-state degeneracy of the system in the cluster limit $\lambda=0$.

Henceforth we shall focus on periodic boundary conditions.

\section{Correlation Functions}
\label{sec:corr_functions}
\begin{figure*}
\begin{center}
\subfigure[\label{fig:corr_x0}]{\includegraphics[width=\columnwidth]{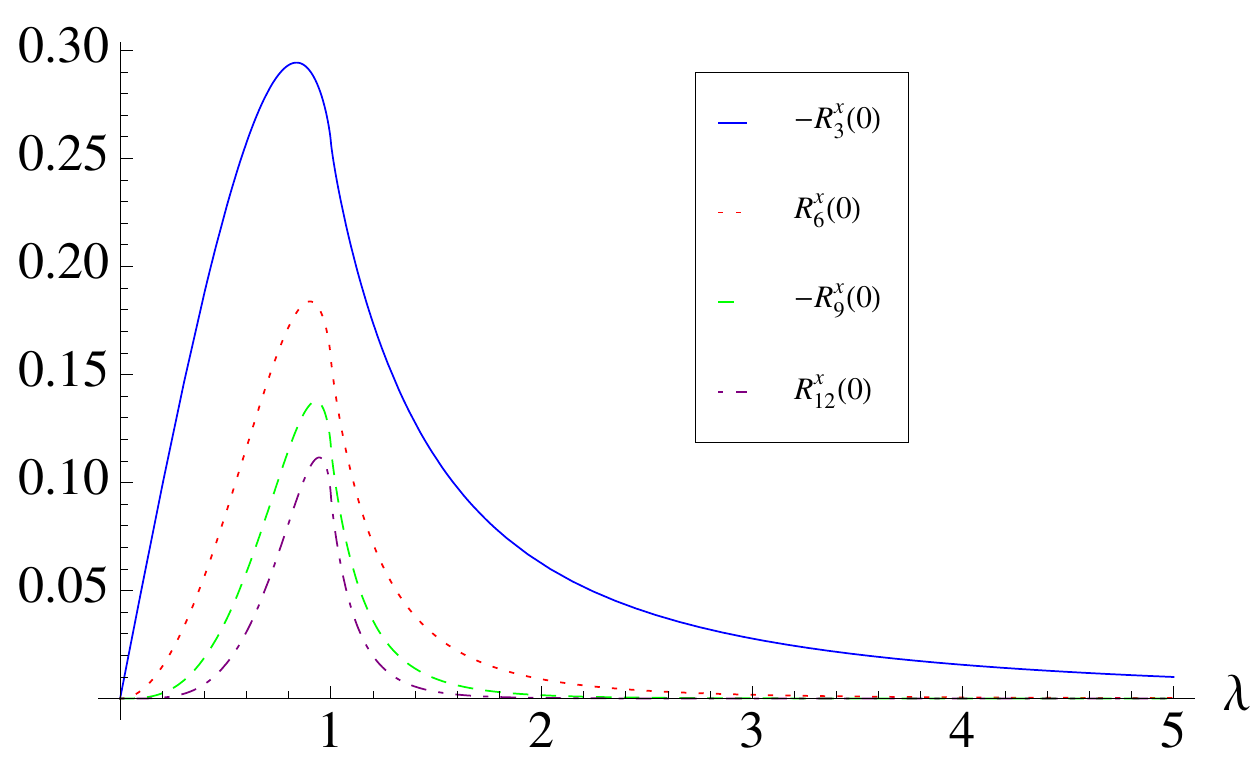}}
\subfigure[\label{fig:corr_xt}]{\includegraphics[width=\columnwidth]{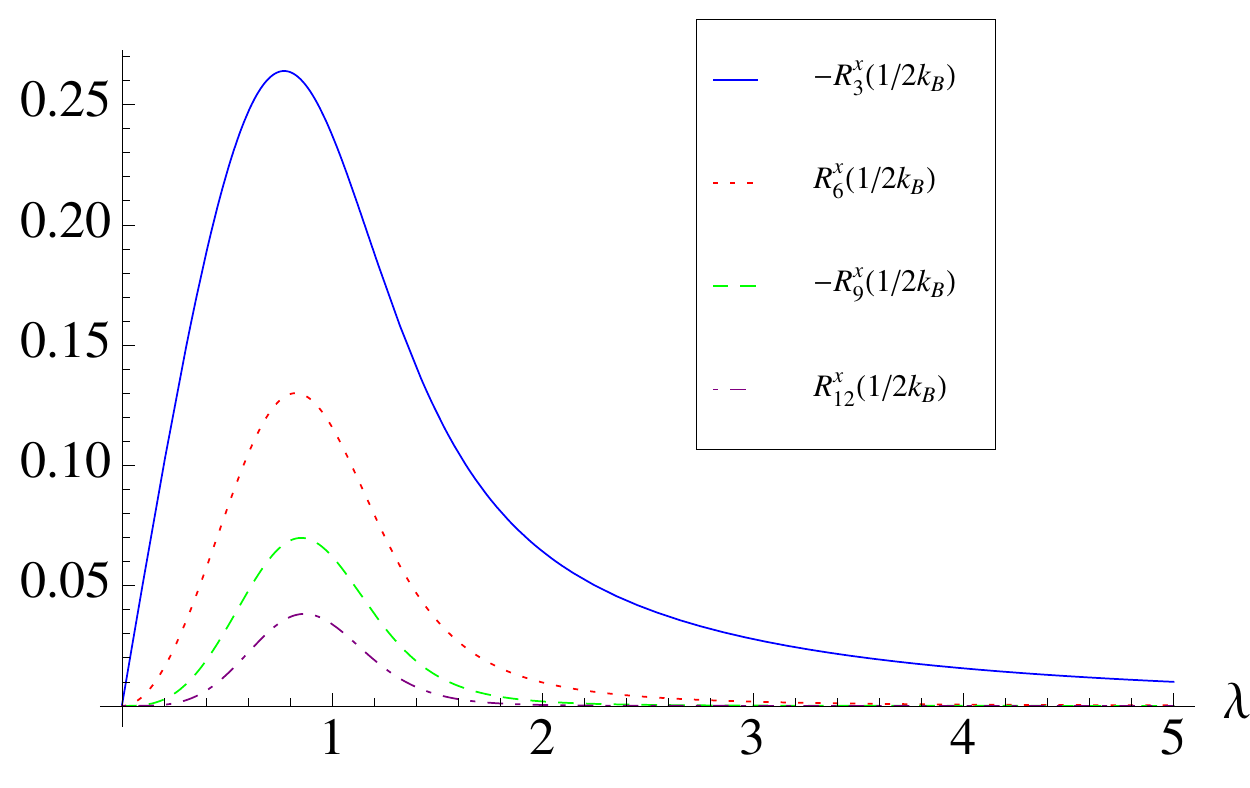}}
\caption{(Color online) Absolute value of the two-point correlation functions at fixed temperature vs $\lambda$: (a) $R_r^x(0)$ for $r=3,6,9,12$; (b) $R_r^x(1/2k_B)$ for $r=3,6,9,12$. $\lambda=1$ is the critical point in the thermodynamic limit.}
\end{center}
\end{figure*}
\begin{figure*}
\begin{center}
\subfigure[\label{fig:corr_y0}]{\includegraphics[width=\columnwidth]{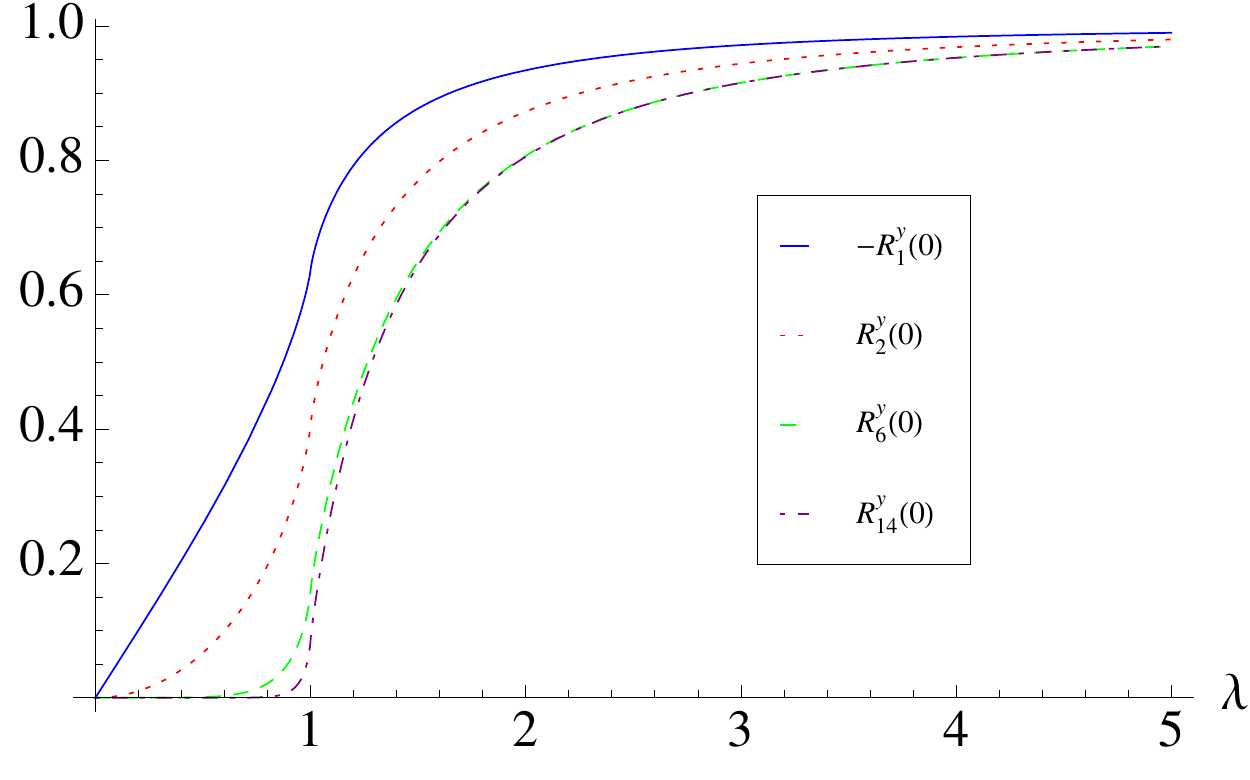}}
\subfigure[\label{fig:corr_yt}]{\includegraphics[width=\columnwidth]{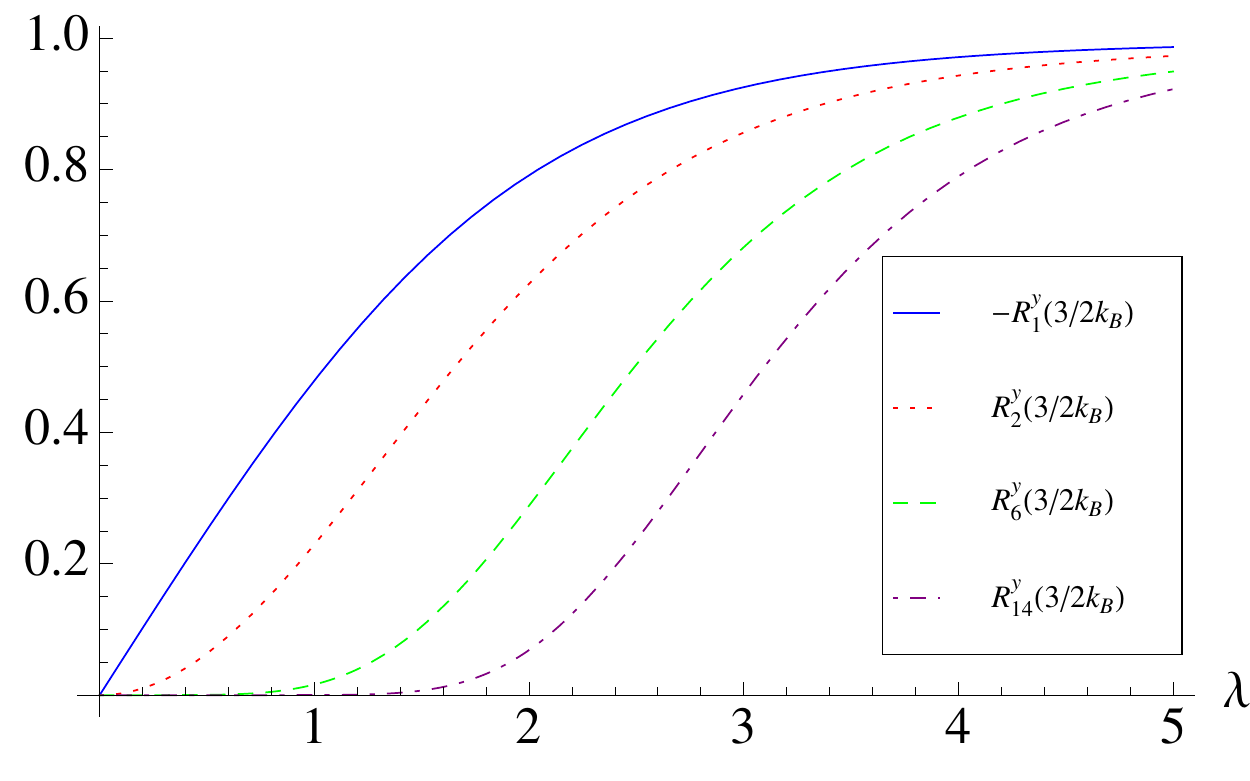}}
\caption{(Color online) Absolute value of the two-points correlation functions at fixed temperature vs $\lambda$: (a) $R_r^y(0)$ for $r=1,2,6,14$; (b) $R_r^y(3/2k_B)$ for $r=1,2,6,14$.  $\lambda=1$ is the critical point in the thermodynamic limit.}
\end{center}
\end{figure*}
In this section we detail  the calculation of the spin correlation functions at temperature $T$, defined as:
\begin{equation}
R_{jl}^\alpha(T)=\langle \sigma_j^\alpha \sigma_l^\alpha\rangle_T \quad \textrm{with} \; \alpha=x,y,z,
\label{eq:correlations_def}
\end{equation}
with 
$\langle \cdot  \rangle_T = \frac{1}{\mathcal{Z}} \textrm{Tr} \left( \cdot \,e^{-H/k_BT}\right)$
denoting expectation values in the canonical  ensemble.
The method we shall employ  is a straightforward application of the techniques adopted in Refs.~\cite{LSM,McCoy71}. 

Let us start considering the correlations $R_{jl}^x (T)$. We find
\begin{eqnarray}
R_{jl}^x (T) & =& {\left\langle \left(c_j-c^{\dagger}_j\right) \prod_{j<m<l} \left(1-2 c_m c^{\dagger}_m\right) \left(c^{\dagger}_l+c_l \right) \right\rangle}_T\nonumber\\
&=&\langle B_j A_{j+1}B_{j+1}\dots A_{l-1}B_{l-1} A_l \rangle_T,
\label{eq:x_corr}
\end{eqnarray}
where 
\begin{equation}
\label{eq:ABdef}
A_j=c^\dagger_j+c_j, \qquad B_j=c_j-c^\dagger_j.
\end{equation}
Using Wick's theorem  we can evaluate  the vacuum expectation value of a product of anticommuting operators  $A_j$'s and $B_j$'s, in term of contractions of pairs thereof. The needed contractions are
\begin{equation} 
\left\{
\begin{aligned}
&\langle A_j A_l \rangle_T =\delta_{jl}\\
&\langle B_j B_l \rangle_T =-\delta_{jl}\\
&\langle B_j A_l \rangle_T =D_{jl}(T)=D(j-l,T)=D(r,T)
\end{aligned}
\right.
\end{equation}
where the function $D$ at a fixed temperature depends only on the relative distance $r=j-l$ between spins, as a consequence of the translational invariance of the system. In the thermodynamic limit the explicit form of this function is
\begin{eqnarray}
D(r,T) &=& \frac{1}{\pi} \int_0^\pi dp\, \frac{\tanh\left(\beta \Lambda(p)\right)}{\Lambda(p)} \nonumber \\
& & \times \left\{\cos \left[(r+2)p\right]-\lambda \cos \left[(r-1)p \right] \right\},
\label{eq:d_function}
\end{eqnarray}
where $\Lambda(p)$ has been defined in Eq.~(\ref{eq:dispersion}).

\begin{figure*}
\begin{center}
\subfigure[\label{fig:corr_x2}]{\includegraphics[width=\columnwidth]{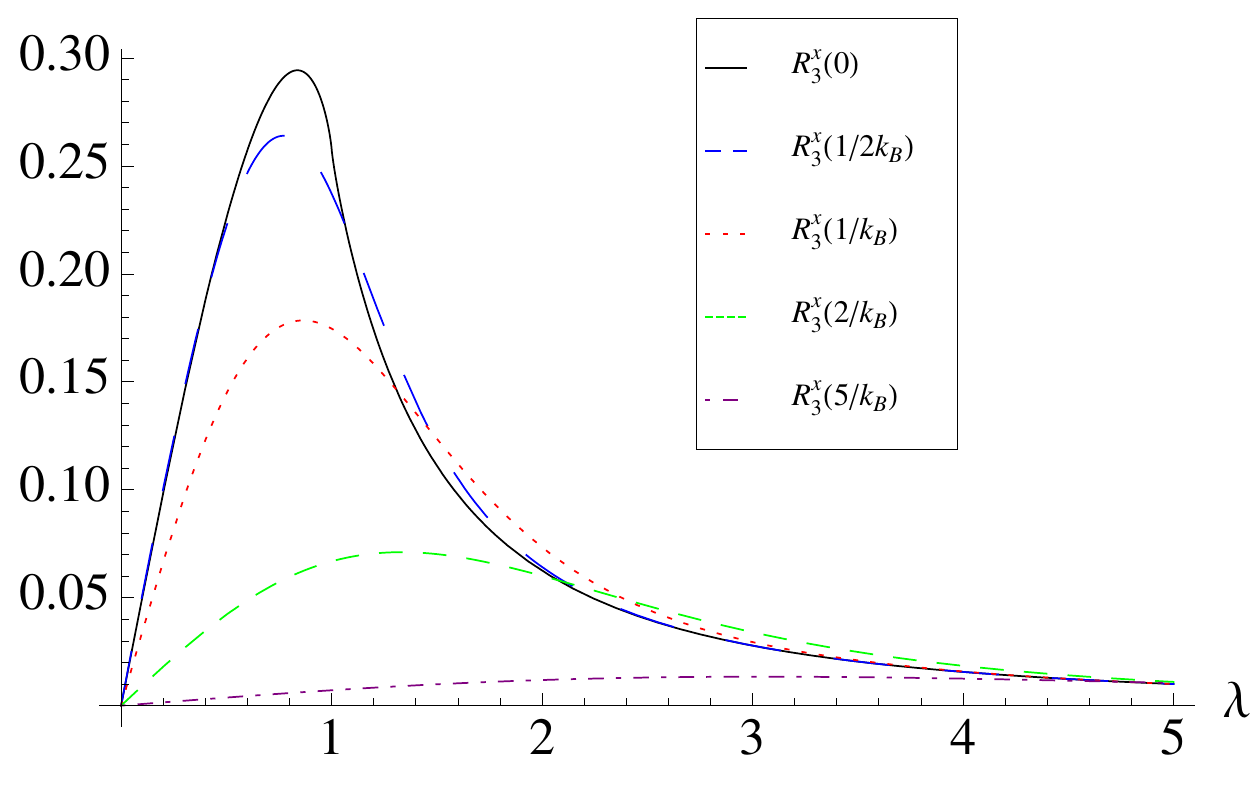}}
\subfigure[\label{fig:corr_y2}]{\includegraphics[width=\columnwidth]{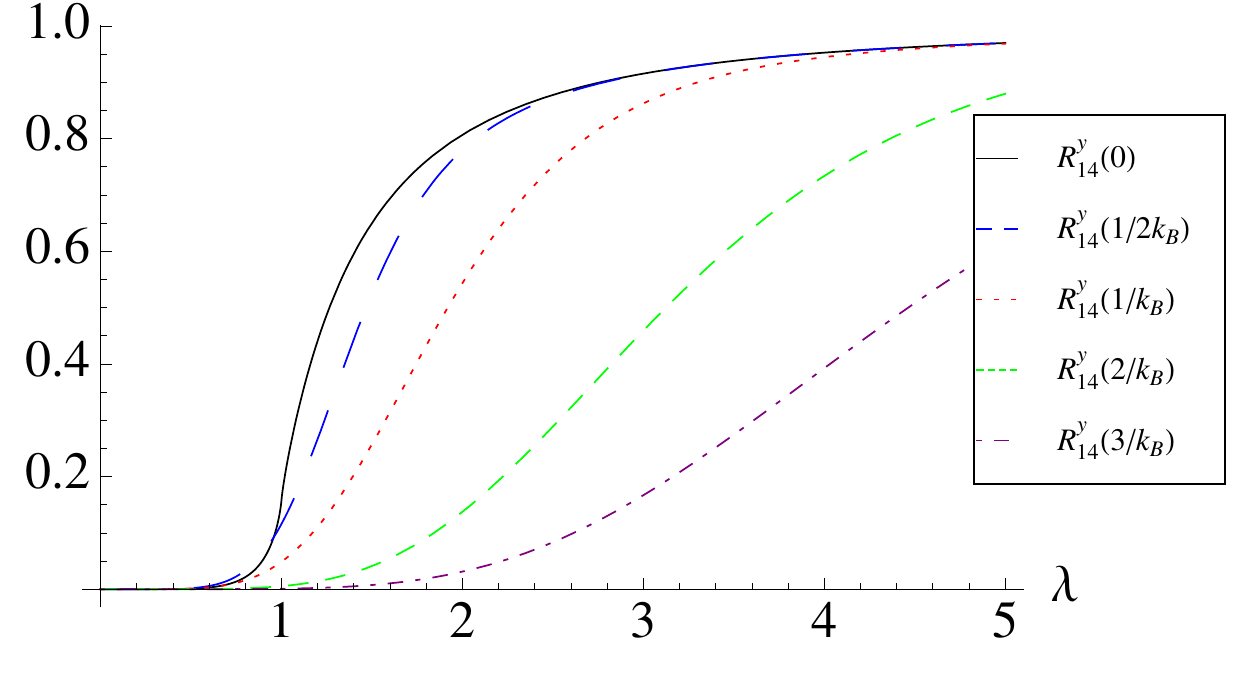}}
\caption{(Color online) Absolute value of the two-points correlation functions at a fixed distance vs $\lambda$: (a) $-R_3^x(T)$ for $T=0,1/2k_B,1/k_B,2/k_B,5/k_B$; (b) $R_{14}^y(T)$ for $T=0,1/2k_B,1/k_B,2/k_B,3/k_B$. 
$\lambda=1$ is the critical point in the thermodynamic limit.}
\end{center}
\end{figure*}

As for the standard Ising model the contractions  $\langle A_j A_j \rangle_T$ and $\langle B_j B_j \rangle_T$  vanish at equilibrium. Therefore the pfaffians can be reduced to determinants.
The resulting spin correlators,  $R_{jl}^x=R_{j-l}^x=R_r^x=R_{-r}^x$, read
\begin{equation}
R_r^x(T)=\begin{vmatrix}
D(-1,T) & D(-2,T) & \cdots &D(-r,T) \\
D(0,T)& D(-1,T) & \cdots &D(-r+1,T) \\
\vdots & \vdots& \ddots& \vdots \\
D(r-2,T) & D(r-3,T) & \cdots & D(-1,T) 
\end{vmatrix}.
\label{eq:x_cluster}
\end{equation}
Similarly, for the other two correlation functions we have
\begin{equation}
R_r^y(T)=\begin{vmatrix}
D(1,T)& D(0,T) & \cdots & D(-r+2,T) \\
D(2,T) & D(1,T) & \cdots & D(-r+3,T)\\
\vdots & \vdots & \ddots & \vdots \\
D(r,T)& D(r-1,T) & \cdots & D(1,T) 
\end{vmatrix},
\label{eq:corr_y}
\end{equation}
and
\begin{equation}
R_r^z(T)= D(0,T)^2 - D(r,T) D(-r,T).
\label{eq:z_corr}
\end{equation}
Finally, we observe that the magnetization along the $z$-axis $m_z$ is given by
\begin{equation}
m_z(T)=\langle \sigma^z_j\rangle_T=\langle 1-2c_jc^\dagger_j\rangle_T=\langle A_j B_j\rangle_T=-D(0,T),
\label{eq:mz}
\end{equation}
where translational invariance has been invoked.
We now show that the system under investigation is endowed with a very specific correlation pattern.
To this end we notice that (\ref{eq:d_function}) has the form
\begin{equation}
D(r,T)= I(r+2)-\lambda I(r-1),
\end{equation} 
where
\begin{eqnarray}
I(n)&=& \frac{1}{\pi}\int_0^\pi dp\,  \cos(np)\, \frac{\tanh\left(\beta \Lambda(p)\right)}{\Lambda(p)} \\
&=&\frac{1}{\pi} \int_0^{\pi/3} dp\, \left[ \cos (np) +2 \cos \left(\frac{2 \pi n}{3}\right) \cos (n p) \right] 
\nonumber\\
& &\times \frac{\tanh\left(\beta \Lambda(p)\right)}{\Lambda(p)}.
\end{eqnarray}
The square bracket, and therefore $I(n)$, is non-vanishing only if $n$ is a multiple of $3$, and, therefore
\begin{equation}
 D(r,T) \neq 0 \iff r= 3m +1, \quad m \in \mathbb{Z}.
 \label{eq:DrT}
 \end{equation}
This result implies that all correlation functions in Eq.~(\ref{eq:z_corr}) and the magnetization 
along the $z$-axis (\ref{eq:mz})  are identically zero. On the other hand, the correlation along the $x$-axis $R_r^x(T)$ is non-vanishing only when $r$ is a multiple of 3, and its absolute value at a fixed temperature decreases when the relative distance increases, as can be seen in Figs.\ \ref{fig:corr_x0} and 
\ref{fig:corr_xt}. Clearly, the $x$-correlation is more sensitive to criticality for smaller $r$.
The correlation along the $y$-axis $R_r^y(T)$ is always non-vanishing and negative (positive) when $r$ is odd (even), as expected because of the anti-ferromagnetic nature of the Ising interaction in~(\ref{eq:ham_cluster}). See Figs.\ \ref{fig:corr_y0} and \ref{fig:corr_yt}.
Its absolute value is small when $\lambda <1$ and tends to its maximum value 1 when $\lambda \to \infty$. Moreover, at zero temperature [Fig.~\ref{fig:corr_y0}], the transition around the critical value $\lambda=1$ becomes more abrupt for larger relative distance $r$, 
while at a nonvanishing temperature [$=3/2k_B$ in Fig.~\ref{fig:corr_yt}] the transition is softer.
Finally, the curves of $R_r^x(T)$ become less and less peaked when distance is kept fixed and temperature is increased.
See Fig.~\ref{fig:corr_x2}.
Analogously, at a fixed distance, the transition to a non-vanishing value of $R_r^y(T)$ is sharper for lower temperatures, as shown in Fig.~\ref{fig:corr_y2}.

The global picture that emerges from the study of the correlation functions is the following. Around criticality, $R_r^x(T)$ is more peaked at small temperatures and small distances $r$; $R_r^y(T)$ undergoes a sharper transition from zero to unity at small temperatures and large distances $r$. Notice the opposite behavior of $R_r^x$ and $R_r^y$ for small/large values of $r$, a consequence of the order in the $y$ direction.

Using these results, in the following subsections we will evaluate the staggered magnetization and the string parameter of our model. We will see that these two quantities are able to capture the nature of the Ising and the cluster phase, respectively, sharply marking the phase transition at $\lambda=1$.

\subsection{Staggered magnetization}
\label{sec:staggered_m}

Because of the $Z_2$ symmetry of the Hamiltonian, we cannot compute the staggered magnetization by directly applying the definition
$m_y=\langle  (-1)^j \sigma^y_j\rangle_T$, as we would always obtain a vanishing result. Hence we use the formula
\begin{equation}
\lim_{r \rightarrow \infty} (-1)^r R_r^y(0) = m^2_y.
\end{equation}
We notice that (\ref{eq:corr_y}), defining $R_r^y$, is a Toeplitz matrix, namely a matrix  $\mathbb{A}$ with  elements $a_{ij}=a_{|i-j|}$.  
Therefore we can apply Szeg\H{o}'s theorem 
enabling us to fix the asymptotics  $r\rightarrow \infty$ of the correlation functions. 
Under the assumption that  
 $a_m$'s can be considered as  the coefficients of a  Fourier series of suitable  function $f(p)$:
$a_m= \frac{1}{2 \pi} \int_{-\pi}^\pi \e^{-\ii m p} f(p) \, dp $, we remind that the theorem states 
that 
\begin{equation}
\lim_{n \rightarrow \infty} \frac{{\rm det}\, \mathbb{A}_n}{\mu^n}=\exp \left(\sum_{n=1}^{\infty} n g_n g_{-n}\right),
\label{eq:szego}
\end{equation}
where
$\log \mu= \frac{1}{2 \pi} \int_{-\pi}^\pi \log f(p) \, dp$,
and
$g_n=\frac{1}{2 \pi} \int_{-\pi}^\pi e^{-\ii n p} \log f(p) \, dp$.
To apply the Szeg\"o theorem to our problem, we write
\begin{equation}
D(r,0)= -\frac{1}{2 \pi} \int_{-\pi}^{\pi} \e^{-\ii r p} \e^{\ii p} \sqrt{\frac{\lambda-\e^{-3 \ii p}}{\lambda-\e^{3\ii p}}} \; dp.
\end{equation}
We consider the integral
\begin{equation}
D(r+1,0)=C(r)=\frac{1}{2 \pi} \int_{-\pi}^\pi \e^{-\ii rp} \tilde{c} (p) \; dp,
\end{equation}
with $\tilde{c} (p) =-\sqrt{\frac{\lambda-\e^{-3\ii p}}{\lambda-\e^{3 \ii p}}}$ 
and 
\begin{equation}
\log \mu = \frac{\ii}{2} \int_{-\pi}^{\pi} dp +\frac{1}{4\pi} \int_{-\pi}^\pi \log \frac{\lambda-\e^{-3 \ii p}}{\lambda-\e^{3\ii p}} \; dp= \ii \pi,
\label{eq:mu_cluster}
\end{equation}
where the second integral vanishes because the integrand is odd. This yields $\mu=-1$.
With this definition we can write the spin correlation function along $y$ as
\begin{equation}
R^y_r(0)= \begin{vmatrix}
C(0) &C(-1) &\cdots & C(-n+1) \\
C(1) &C(0)  &\cdots & C(-n+2)\\
\vdots & \vdots  & \ddots & \vdots\\
C(n-1) & C(n-2) & \cdots &C(0).
\end{vmatrix}.
\end{equation}

In order to compute the $g_n$'s, we start from the case $\lambda>1$. Defining $\alpha=1/\lambda$, we obtain:
\begin{equation}
\log \tilde{c} (p) =\ii \pi + \ii \sum_{l=1}^{\infty} \frac{\alpha^l}{l} \sin(3 lp).
\end{equation}
As a consequence we have
\begin{eqnarray}
g_n&=&\frac{1}{2 \pi} \sum_{l=1}^{\infty} \frac{\alpha^l}{l} \int_{-\pi}^\pi \sin(np) \sin (3 l p)\;dp \nonumber\\
& =& \begin{cases} \frac{3}{2n} \alpha^{n/3} & \text{if $n\in B=\{3,6,9,\dots\}$} \\
0 & \text{otherwise}
\end{cases}
\end{eqnarray}
and
\begin{equation}
g_{-n}= -g_n =-\frac{3}{2n} \alpha^{n/3}.
\end{equation}
Finally,
\begin{equation}
\sum_{n=1}^\infty n g_n g_{-n}=-\frac{9}{4} \sum_{n \in B} \frac{\alpha^{2n/3}}{n} =-\frac{3}{4} \log \left(1-\alpha^2\right).
\label{eq:g_cluster}
\end{equation}
When $\lambda<1$ one can proceed in an analogous way, obtaining
\begin{equation}
\sum_{n=1}^n n g_{n} g_{-n} =-\infty.
\label{eq:g_cluster2}
\end{equation}
By plugging Eqs.~(\ref{eq:mu_cluster}), (\ref{eq:g_cluster}), and (\ref{eq:g_cluster2}) into (\ref{eq:szego}) and (\ref{eq:corr_y}) we obtain 
\begin{eqnarray}
 m_y &=& \pm\sqrt{\lim_{r\rightarrow \infty} (-1)^r R_r^y(0)} \nonumber \\
&=& \begin{cases} \pm \left(1-\lambda^{-2}\right)^{3/8} & \text{when $\lambda>1$} \\
0 & \text{when $\lambda<1$}
\end{cases}.
\label{eq:val_my}
\end{eqnarray}
Therefore the staggered magnetization along the $y$-axis captures the antiferromagnetic order in the Ising phase $\lambda>1$. The quantity $m_y$ is displayed in Fig.~\ref{fig:order_parameters}. It plays the role of order parameter for $\lambda >1$, but it is nonlocal.

\subsection{String correlation functions}
\label{sect:string}
We  compute here the string order parameter 
\begin{equation}
O_z = \lim_{N\to\infty}(-1)^N \left\langle \sigma_1^x \sigma_2^y \left(\prod_{k=3}^{N-2} \sigma_k^z\right) \sigma_{N-1}^y \sigma_{N}^x \right\rangle_0 ,
\label{eq:Oz}
\end{equation}
by exploiting the duality of the model discussed in Sec.~\ref{sect:duality}.
The expression of the dual variables $\mu_j^x$ and $\mu_j^z$ in Eq.~(\ref{eq:dual}) implies 
\begin{equation}
\mu_j^y= -\ii \sigma_j^z \sigma_j^x = - \left(\prod_{k=1}^{j-1} \sigma_k^z\right) \sigma_j^y \sigma_{j+1}^x.
\end{equation}
We consider the dual $y$-correlation function between site $j$ and site $l$
\begin{equation}
\mu_j^y \mu_{l}^y=\sigma_j^x \sigma_{j+1}^y \left(\prod_{k=j+2}^{l-1} \sigma_k^z\right) \sigma_{l}^y \sigma_{l+1}^x,
\end{equation}
which for $j=1$ and $l=N-1$ particularizes to
\begin{equation}
\mu_1^y \mu_{N-1}^y=\sigma_1^x \sigma_2^y \left(\prod_{k=3}^{N-2} \sigma_k^z\right) \sigma_{N-1}^y \sigma_{N}^x.
\end{equation}
In the    thermodynamic limit $N\rightarrow \infty $ we can apply  Szeg\H{o}'s theorem and the string correlation can be calculated in terms of the staggered magnetization:
\begin{equation}
O_z = \lim_{N\rightarrow\infty} (-1)^{N}\langle\mu_1^y \mu_{N-1}^y \rangle_0 =m^2_{y, \rm{dual}}\; .
\end{equation}

Because of the duality in Eq.~(\ref{eq:dual2}), the quantity $m^2_{y, \rm{dual}}$ can be calculated with the procedure followed in Section~\ref{sec:staggered_m} and is given by~(\ref{eq:val_my}) after replacing $\lambda$ with $1/\lambda$, namely
\begin{equation}
O_z= m^2_{y, \rm{dual}} =\begin{cases}  \left(1-\lambda^2\right)^{3/4} & \text{when $\lambda<1$} \\
0 & \text{when $\lambda>1$}
\end{cases}.
\label{eq:string_order}
\end{equation}
This is the quantity plotted in Fig.~\ref{fig:order_parameters}.
It is nonvanishing for $\lambda<1$ and captures the hidden order of the cluster phase.
\subsection{Finite Size Correlations Functions}
\label{sec:finite_corr}

\begin{figure*}
\begin{center}
\subfigure[ \label{fig:corrx_finite}]{
\includegraphics[width=\columnwidth]{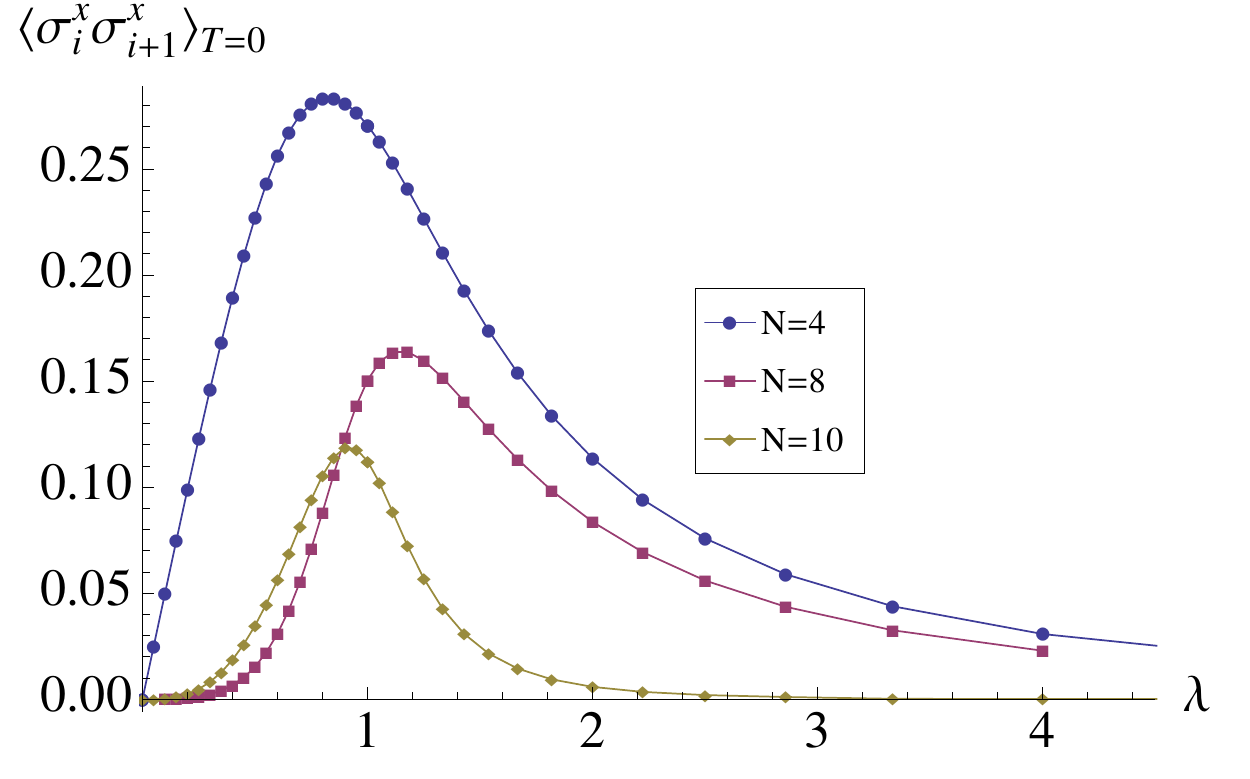}}
\subfigure[\label{fig:magnz_finite}]{
\includegraphics[width=\columnwidth]{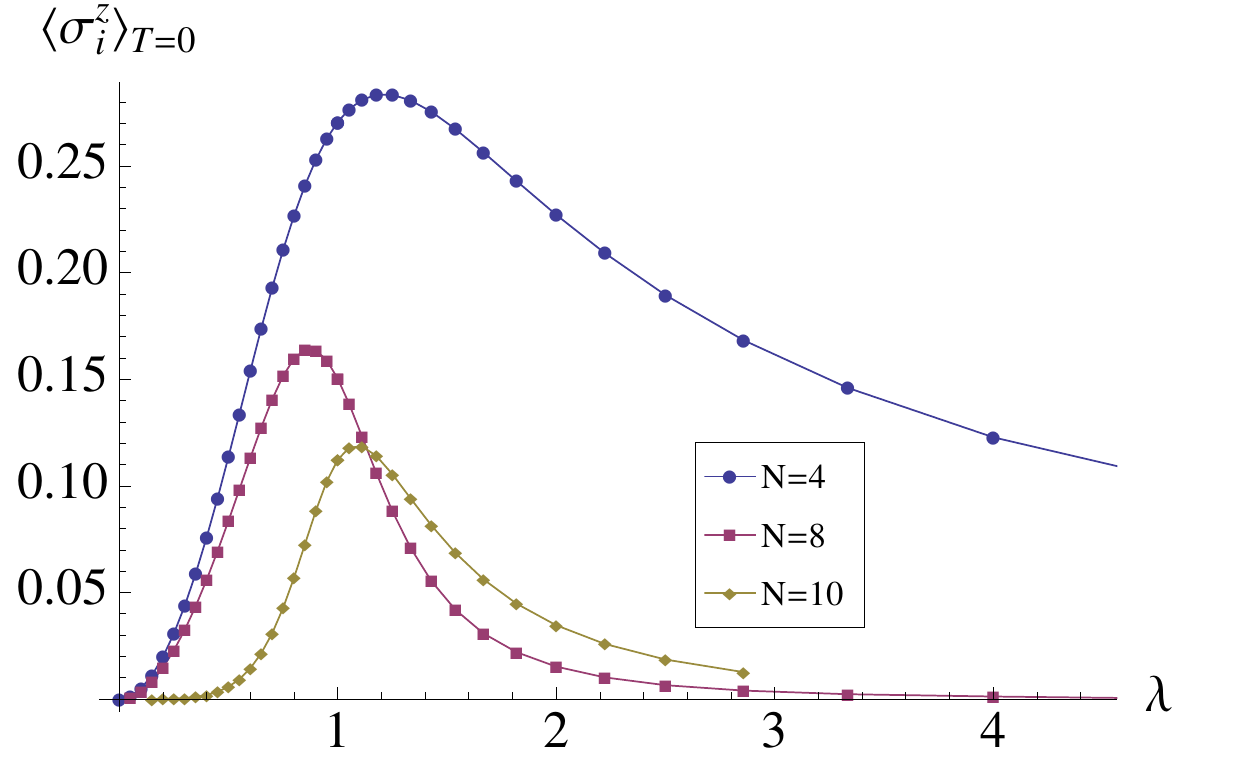}}
\caption{(Color online) (a) First-neighbor zero-temperature two-points correlation function along the $x$-axis vs $\lambda$. (b) Zero-temperature magnetization along the $z$-axis vs $\lambda$. In both figures,  $N=4,8$ and $10$.
}
\label{fig:corr_finite}
\end{center}
\end{figure*}

We conclude our analysis of correlations by briefly looking at the behavior of the correlation functions and magnetization along the $z$-axes at zero temperature, when the number of sites $N$ is finite. We shall use numerical methods, starting from the Hamiltonian (\ref{eq:ham_cluster}), considering the boundary terms. 
In particular, we consider the cases $N=4,6,8,10$ and $12$, with an exact diagonalization of the Hamiltonian and computing the expectation values (\ref{eq:correlations_def}) on the ground state of the system. We limit ourselves to the case of first neighbor spins.

In contrast with the thermodynamic limit [see comments following Eq.~(\ref{eq:DrT})], the correlations along $x$ and $z$ and the magnetization along $z$ are in general non-vanishing. However, they do vanish when the number of sites is a multiple of 3, that is when $N=6$ and $N=12$
[see again comments following Eq.~(\ref{eq:DrT})].
The absolute values of these ``spurious'' correlations decrease when the size of the system is increased. In particular, their maximum vanishes like $0.97 \times N^{-0.88}$, in accord with the thermodynamic limit.
The behavior of the correlation along the $x$-axis and of the magnetization along the $z$-axis is shown in Fig.~\ref{fig:corr_finite} (the plots of the correlation along $z$ being analogous). This finite-size analysis corroborates the findings of the present section.

\subsection{Correlations between Majorana operators}
\label{majorana_block}
We study here the ground-state correlations between Majorana operators, by employing 
the Majorana representation of the system, introduced in Eq.~(\ref{majorana_fermions}).
From Eq.~(\ref{eq:ABdef}) we get $\check{a}_{2j-1}=A_j$ and $\check{a}_{2j}= \ii B_j$, so that
\begin{equation} 
\label{eq:majorana_corr}
\left\{
\begin{aligned}
&\langle \check{a}_{2j-1} \check{a}_{2l-1} \rangle_0=\langle A_j A_l \rangle_0 =\delta_{jl}\\
&\langle \check{a}_{2j-1} \check{a}_{2l} \rangle_0=\ii \langle A_j B_l \rangle_0 = -\ii D(r,0)\\
&\langle \check{a}_{2j} \check{a}_{2l-1} \rangle_0=\ii \langle B_j A_l \rangle_0 =\ii D(-r,0)\\
&\langle \check{a}_{2j} \check{a}_{2l} \rangle_0=-\langle B_j B_l \rangle_0=\delta_{jl}
\end{aligned}
\right.
\end{equation}
where $r=j-l$ and $D(r)$ is defined in Eq.~(\ref{eq:d_function}). Equations~(\ref{eq:majorana_corr}) can be summarized as:
\begin{equation}
\label{eq:deltagamma}
\langle \check{a}_{j} \check{a}_{l} \rangle_{0} =\delta_{jl}+\ii (\Gamma_N)_{jl} ,  \quad  j,l=1,\dots, 2 N,
\end{equation}
where
\begin{eqnarray}
\Gamma_N&=&\begin{bmatrix}
\Pi_0 & \Pi_{-1} & \dots & \Pi_{-N+1}\\
\Pi_1 & \Pi_0 & \dots & \Pi_{-N+2} \\
\vdots & \vdots & \ddots & \vdots \\
\Pi_{N-1} & \Pi_{N-2} &\dots & \Pi_0
\end{bmatrix}, \\
\Pi_j &=&\begin{bmatrix} 
0 & D(j,0) \\
-D(-j,0) & 0
\end{bmatrix} 
\end{eqnarray}
with $D(r,0)$ defined in Eq.~(\ref{eq:d_function}).
These expressions will be useful in the following, in particular in Secs.~\ref{sec:reddens} and~\ref{sec:entropy}.

\section{Reduced density operators}
\label{sec:reddens}

We provide here explicit expressions of the reduced density matrix  for a single spin, two spins and a block of  $L$ spins.
Let us briefly derive the reduced spin density matrices by suitable partial tracing of the state $\rho$ of the spin system: 
\begin{eqnarray}
\rho_{j_1\dots j_L}& =& \sum_{\alpha_{j_{L+1}},\dots, \alpha_{j_N}}  \sum_w p^w |\alpha_{j_1}\dots \alpha_{j_L} \rangle \langle \alpha_{j_1}\dots \alpha_{j_L} | \nonumber 
\\ 
&& \langle\alpha_{j_{L+1}}\dots \alpha_{j_N} |\Omega_w\rangle \langle \Omega_w| \alpha_{j_{L+1}}\dots \alpha_{j_N}\rangle .
\end{eqnarray}
where $|\Omega_w\rangle$ denote the states of the chain arising with probability $p^w$ and $\alpha\in\{\uparrow,\downarrow\}$. Here $(j_1,\dots, j_N)$ is a given permutation of $(1,\dots,N)$.  

First we consider the  reduced density matrix of a block of contiguous spins; the single and two spin cases will be obtained as particular examples.
By translational invariance we can consider, without loss of generality, the block of the first $L$ spins. Then we can expand the matrix $\rho_{\{L\}}=\rho_{1 2\dots L}$ as
\begin{equation}
\rho_{\{L\}}= 2^{-L} \sum_{\alpha_1,\dots, \alpha_L \in\{0,x,y,z\}} p_{\alpha_1 \dots \alpha_L} \sigma_1^{\alpha_1}\dots \sigma_L^{\alpha_L},
\label{eq:block_expansion}
\end{equation}
where
\begin{equation}
p_{\alpha_1 \dots \alpha_L}=\langle \sigma_1^{\alpha_1} \dots \sigma_L^{\alpha_L}\rangle
\end{equation}
can be obtained through the expressions of the Majorana operators outlined in Section \ref{majorana_block}. 
In fact, we know that the system is invariant under parity transformation, hence $p_{\alpha_1 \dots \alpha_L}=0$ whenever the sum of the $\alpha$'s equal to $x$ and of the $\alpha$'s equal to $y$ is odd. Therefore the non-vanishing coefficients of the expansion (\ref{eq:block_expansion}) correspond to the expectation values of a product of Pauli matrices with an even total number of $\sigma_x$'s and $\sigma_y$'s. Products of this kind are mapped into products of an even number of Majorana operators $\check{a}_j$, with $j = 1,\dots, 2L$. We can then can use the Wick theorem to express such products in terms of the correlations $\langle \check{a}_j \check{a}_l \rangle$ 
which are contained in the matrix $\Gamma_N$ of Eq.~(\ref{eq:deltagamma}).

Let us now consider the state of a single spin placed at site $j$ of the chain. 
In general, the spin will be in a mixed state.
The Hilbert space  is $\mathbb{C}^2$ and in the basis $\{\sigma_j^0=I_j,\sigma_j^x,\sigma_j^y,\sigma_j^z\}$  the single-spin reduced density matrix is given by
\begin{equation}
\label{eq:reduced1}
\rho_j=\frac{1}{2} \sum_{\alpha\in\{0,x,y,z\}} p_{\alpha} \sigma_j^\alpha,
\end{equation}
where
$p_\alpha= {\rm Tr} \left(\sigma^\alpha_j \rho_j \right)= \langle \sigma_j^\alpha \rangle$.
The system considered in this article is traslationally invariant, and therefore the single-spin state, that we denote $\rho_1$, is translationally invariant as well.

Finally,   for the case of two spins placed at sites $j$ and $l$ of the chain,
the matrix $\rho_{jl}$ acts on the Hilbert space $\mathbb{C}^2 \otimes \mathbb{C}^2$ of which the set $\left\{\sigma_j ^\alpha \otimes \sigma_l^ \beta\right\}$, with $\alpha,\beta\in\{0,x,y,z\}$, is an orthonormal basis. Therefore  we can make a formal expansion of the reduced density operator on such a basis 
\begin{equation}
\rho_{jl}=\frac{1}{4} \sum_{\alpha, \beta \in\{0,x,y,z\}} p_{\alpha \beta}\, \sigma_j^{\alpha} \otimes \sigma_l^{\beta},
\label{eq:dec_rho}
\end{equation}
where
$p_{\alpha \beta}= {\rm Tr} \left( \sigma_j^{\alpha} \sigma_l^{\beta}\rho_{jl}\right)= \langle \sigma_j^{\alpha} \sigma_l^{\beta} \rangle$.
Equation (\ref{eq:dec_rho}) makes it manifest that the state of the couple of spins $j$ and $l$ can be written in terms of the spin correlation functions computed in Sec.~\ref{sec:corr_functions}.

\section{Spin entanglement in the ground state}
\label{sec:entanglement}

In this section we analyze the entanglement between two spins and between a single spin and the rest of the system. As measures of such quantity we will make use of \emph{concurrence}  \cite{concurrence} and \emph{residual entanglement}  \cite{residual} respectively. The concurrence is defined as 
\begin{eqnarray}
C(\rho_{1,r+1})&=&C(r)\nonumber\\
&=&{\rm max} \left\{\sqrt{\gamma_1}-\sqrt{\gamma_2}-\sqrt{\gamma_3}-\sqrt{\gamma_4},0\right\},\nonumber\\
\end{eqnarray}
where $\gamma_1\geq\gamma_2\geq\gamma_3\geq\gamma_4\geq 0$ are the eigenvalues  of the matrix $\mathcal{R}_r=\rho_{1,r+1} \tilde{\rho}_{1,r+1}$, with $\tilde{\rho}_{1,r+1}=\left(\sigma^y\otimes \sigma^y\right) \rho^*_{1,r+1} \left(\sigma^y\otimes \sigma^y\right)$, where  $\rho_{1,r+1}$ is the reduced density matrix~(\ref{eq:dec_rho}) and the complex conjugation is taken in the computational basis. Concurrence varies from $C=0$ for a separable state to $C=1$ for a maximally entangled state
and is a convex functional:
\begin{equation}
\label{eq:convexity}
C\left(\sum_i p_i \rho_i \right) \leq \sum_i p_i C(\rho_i).
\end{equation}
The residual entanglement (tangle) is defined as 
\begin{equation}
\tau
=4 \, {\rm det} \rho_1-\sum_{r=1}^{N-1} C^2(r)
\label{eq:tanglej1}
\end{equation}
where $\rho_1$ is the 1-spin reduced density matrix~(\ref{eq:reduced1}). 
The tangle is a global measure of the multipartite entanglement encoded in a (pure) state.

\subsection{Entanglement of the ``thermal'' ground state}
\label{sec:ent_thermal}

\begin{figure*}
\begin{center}
\subfigure[ \label{fig:eigenvalues_3}]{
\includegraphics[width=\columnwidth]{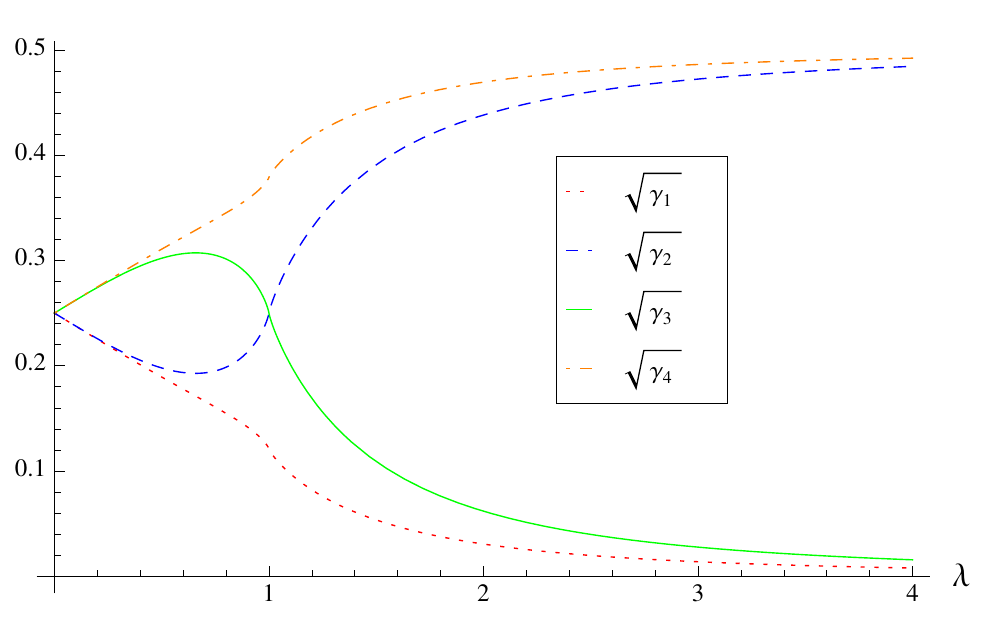}}
\subfigure[\label{fig:eigenvalues_6}]{
\includegraphics[width=\columnwidth]{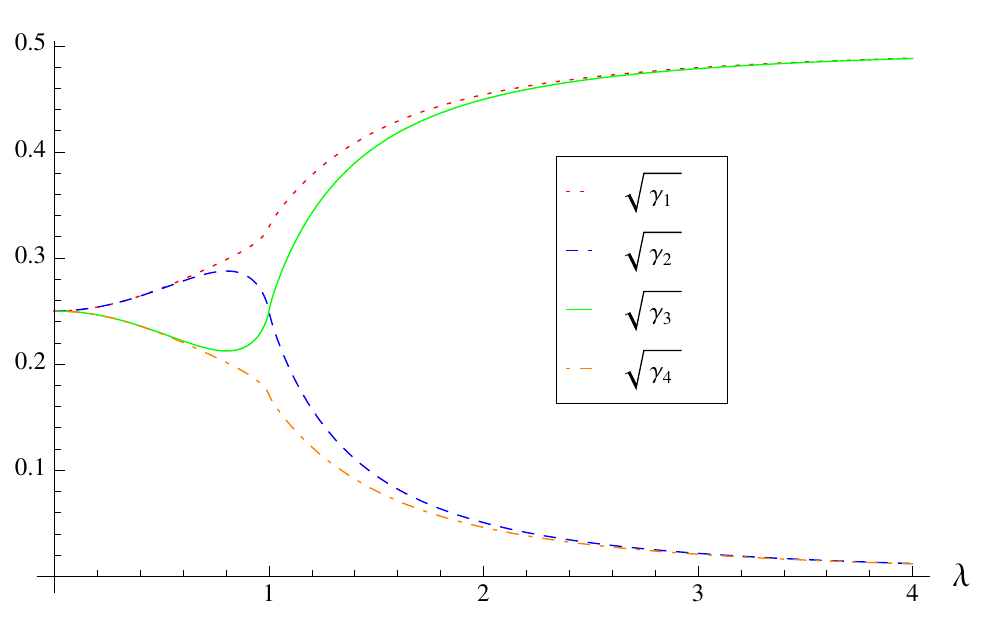}}
\caption{(Color online) Square roots of the eigenvalues $\gamma_i$'s of the matrix $\mathcal{R}_r=\rho_{1,r+1} \tilde{\rho}_{1,r+1}$ vs $\lambda$, when (a) $r=3$ and (b) $r=6$.}
\label{fig:eigenvalues}
\end{center}
\end{figure*}
Let us consider the concurrence between couples of spins when the system is in the ``thermal'' (completely mixed) ground state $\rho_0$ arising as the limit $\beta \rightarrow \infty$ of the thermal canonical state $\rho=\left(1/\mathcal{Z}\right) e^{-\beta H}$: 
\begin{equation}
\rho_0= \frac{1}{2} \left(\ket{\Omega_1} \bra{\Omega_1}+\ket{\Omega_2} \bra{\Omega_2}\right),
\label{eq:thermal_gs}
\end{equation}
where $\ket{\Omega_{1,2}}$ are the two degenerate ground states (we remind that we are considering periodic boundary conditions).
By construction, $\rho_0$ has the same symmetries as the Hamiltonian and, therefore, cannot spontaneously break any symmetry.

A straightforward computation yields the following eigenvalues of the matrix $\mathcal{R}_r$
\begin{equation} 
\label{eq:T01}
\left\{
\begin{aligned}
&\gamma_1= \frac{1}{16}\left(1+R_r^x(0)+R_r^y(0)\right)^2\\
&\gamma_2= \frac{1}{16}\left(1+R_r^x(0)-R_r^y(0)\right)^2\\
&\gamma_3= \frac{1}{16}\left(1-R_r^x(0)+R_r^y(0)\right)^2\\
&\gamma_4= \frac{1}{16}\left(1-R_r^x(0)-R_r^y(0)\right)^2
\end{aligned}
\right.
\end{equation}
We notice that, when $r$ is not a multiple of $3$, the correlation $R_r^x$ vanishes, so that the eigenvalues become pairwise equal:
\begin{equation} 
\left\{
\begin{aligned}
&\gamma_1=\gamma_3=\frac{1}{16} \left(1+R_r^y(0)\right)^2\\
&\gamma_2=\gamma_4=\frac{1}{16}\left(1-R_r^y(0)\right)^2
\end{aligned}
\right.
\end{equation}
This implies that the concurrence $C(r)$ vanishes when $r$ is not a multiple of $3$. On the other hand, when $r$ is a multiple of $3$ the four eigenvalues are in general distinct from each other but the difference of their square roots remains negative, making the concurrence vanish also in this case. Figure \ref{fig:eigenvalues} shows the behavior of the eigenvalues for $r=3$ and $r=6$.

\begin{figure}[h]
\begin{center}
\includegraphics[width=\columnwidth]{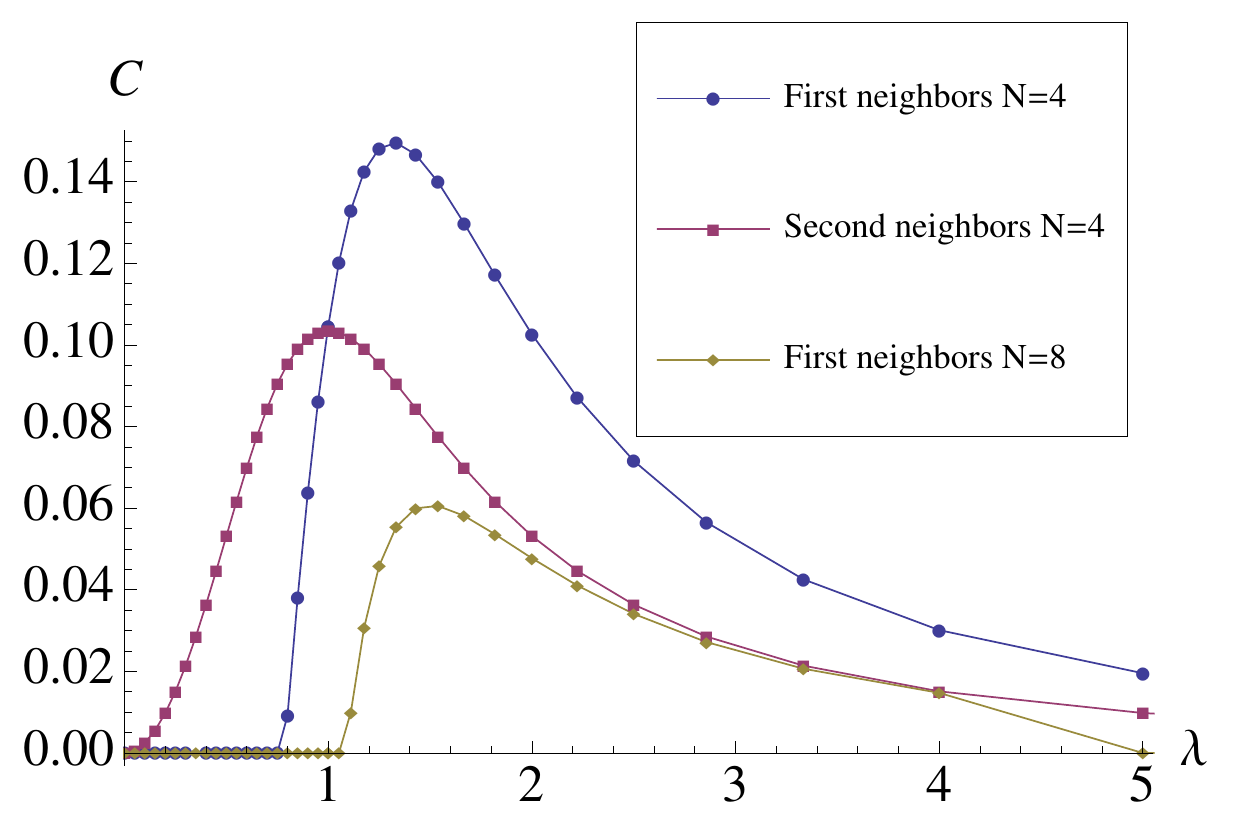}
\caption{(Color online) Non-vanishing concurrences vs $\lambda$ in the ``thermal'' ground state of finite-size systems.}
\label{fig:conc_finite}
\end{center}
\end{figure}

It is instructive to see how the concurrence vanishes in systems of small size. 
We numerically analyzed this problem and our main results are shown in Fig.~\ref{fig:conc_finite}.
In the thermal case, when $N=4$, the concurrence is non-vanishing, both for first and second neighbors; when $N=6$ the concurrence always vanishes, while for $N=8$ only the first-neighbor concurrence is non-vanishing, with a value that is much smaller than that for $N=4$; finally for $N=10$ and $N=12$ the concurrence always vanishes, as in the thermodynamic limit. 
The particular behavior for $N=6$ can be viewed as a consequence of the fact that the correlations have the same value as in the thermodynamic limit, as seen in Sec.~\ref{sec:finite_corr}.
For $N=4$ and $8$ we have finite-size effects similar to those seen in the case of the correlation functions. The case $N=10$ is particularly interesting: even though the correlation functions behave differently from their thermodynamic limit
(see Fig.~\ref{fig:corr_finite}), concurrence identically vanishes. The thermodynamic limit appears therefore to be reached earlier in terms of concurrence than in terms of correlation functions.

In conclusion, for the ``thermal'' ground state (\ref{eq:thermal_gs}),
\begin{equation}
C(r)=0 \quad \forall r. \quad (T=0)
\label{eq:concurrence}
\end{equation}
We conclude that in the ``thermal'' ground state there is no entanglement between any pairs of spins, at any distance from each other, for \emph{any value of $\lambda$}. The concurrence of the ``thermal'' ground state is therefore unable to signal the phase transition at $\lambda=1$.

Incidentally, we observe that the same analysis can be applied to the canonical thermal states at finite temperatures, by replacing $R_r^\alpha(0)$ with $R_r^\alpha(T)$  in Eqs.~(\ref{eq:T01}).
The eigenvalues $\gamma_i$'s have the same properties reported above and therefore, as was to be expected, the result (\ref{eq:concurrence}) is valid at any temperature
\begin{equation}
C(r)=0 \quad \forall r. \quad (T \neq 0)
\label{eq:concurrenceT}
\end{equation}

Finally, let us compute the residual (multipartite) entanglement $\tau$, defined in Eq.~(\ref{eq:tanglej1}), when the global system is in the ``thermal'' ground state (\ref{eq:thermal_gs}). 
One obtains
\begin{equation}
\tau=1, \quad \forall \lambda.
\label{eq:tangleequal1}
\end{equation}
Eqs.~(\ref{eq:concurrence}) and (\ref{eq:tangleequal1}) enable us to conclude that neither the two-spin nor the multipartite entanglement of the ``thermal'' ground state are able to detect the quantum phase transition. A natural question arises: does the  entanglement of the symmetry-breaking ground state detect the phase transition? This problem will be tackled below.

\subsection{Entanglement of the ground state with broken symmetry}

We now discuss the effect of spontaneous symmetry breaking on the ground-state entanglement.  Symmetry breaking is achieved by adding to the Hamiltonian a little staggered magnetic field along the $y$-axis  $h \sum_j (-1)^j \sigma_j^y$, that breaks the invariance of the Hamiltonian under parity transformation. We start with two-spin entanglement (concurrence), by performing a numerical analysis on small systems, like in Sec.~\ref{sec:finite_corr}. 

We start by observing that the convexity of the concurrence, Eq.~(\ref{eq:convexity}), when applied to the ``thermal'' ground state (\ref{eq:thermal_gs}) yields
\begin{equation}
0=C\left(\rho_0 \right) \leq \frac{1}{2}\left[C\left(\rho_1\right)+
C\left(\rho_2\right)\right] =  C\left(\rho_1\right),
\label{eq:conc_ineq}
\end{equation}
due to (\ref{eq:concurrence}) and the symmetry between $\rho_1=\ket{\Omega_1}\bra{\Omega_1}$ and $\rho_2=\ket{\Omega_2}\bra{\Omega_2}$.
Therefore, in principle, the concurrence of the symmetry-breaking ground state could be nonvanishing (and possibly detect the phase transition). We now show that this does \emph{not} happen.

\begin{figure}
\begin{center}
\includegraphics[width=\columnwidth]{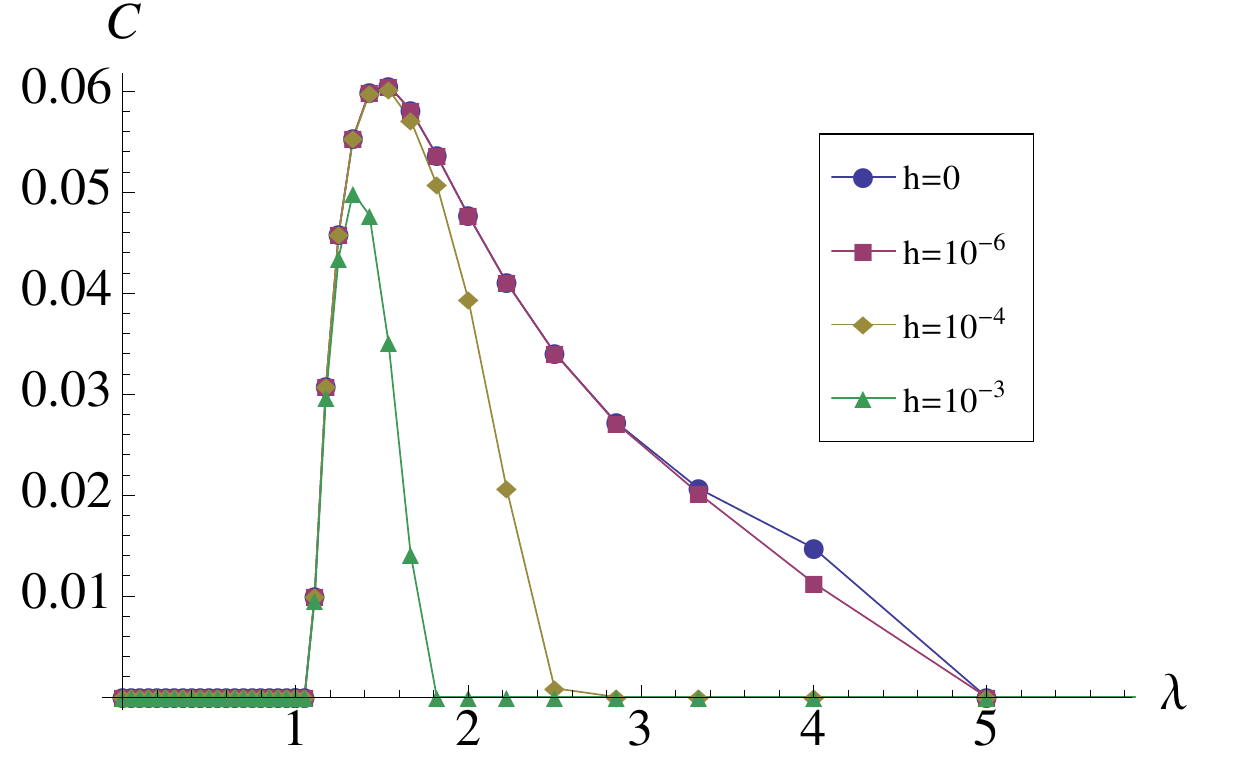}
\caption{(Color online) First-neighbor concurrence vs $\lambda$, for different values of the intensity $h$ of the symmetry-breaking staggered magnetic field and $N=8$.
The points for $h=0$ are taken from Fig.~\ref{fig:conc_finite} ($N=8$).
}
\label{fig:conc_rotta}
\end{center}
\end{figure}

Figure \ref{fig:conc_rotta} displays the first-neighbor concurrence for the symmetry-breaking ground state of a finite chain of $N=8$ spins and a varying strength of the staggered magnetic field.
The plot indicates that concurrence is negligibly affected by the symmetry breaking mechanism: indeed, for $h \to 0$ the curve becomes identical to that of Fig.~\ref{fig:conc_finite}  at $N=8$ (approximating it from below).
The concurrence of second and third neighbors, as well as for longer chains ($N>10$) yields identical results.
This is an evidence that 
there is no bipartite entanglement between couples of spins of the chain, even when one of the degenerate ground states is chosen, breaking the parity symmetry. 
Therefore we conclude that two-spin entanglement is unable to signal the phase transition.

We finally analyze the effect of symmetry breaking on the residual entanglement of the ground state. To this end, we compute the state of spin $j$ when one of the degenerate ground states is chosen.

The expansion of Sec.~\ref{sec:ent_thermal} is still valid, but the values of the coefficients $p_{\alpha}$ [that define the state, see Eq.~(\ref{eq:reduced1})] are modified. The quantities that are symmetric under parity transformation are unaltered, and therefore the  magnetization along the $z$-axis is still vanishing. On the other hand, the values of $\langle \sigma_j^x \rangle$ and $\langle \sigma_j^y \rangle$ can be finite. 
With a calculation very similar to that presented in Sec.~\ref{sec:staggered_m}, one easily shows that 
\begin{equation}
\langle \sigma_j^x \rangle_0^2=\lim_{r \rightarrow \infty} R_r^x(0)
\end{equation}
vanishes for any value of $\lambda$. Therefore the reduced density matrix of spin $j$ is
\begin{equation}
\rho_j =\frac{1}{2} \left[{I}_j+ (-1)^j m_y \sigma_j^y \right].
\end{equation}
On the other hand  the concurrence vanishes and therefore the residual entanglement is
\begin{equation}
\tau= 4{\rm det} \rho_j = 1- m^2_y= \begin{cases}
1 & \text{if $\lambda <1$}\\
1- \left(1-\lambda^{-2}\right)^{3/4} & \text{if $\lambda>1$}
\end{cases}.
\end{equation}
This is shown in Fig.~\ref{fig:ent_res}.
This result indicates that the multipartite entanglement of the ground state with broken symmetry saturates in the cluster phase ($\lambda<1$), whereas it decreases in the antiferromagnetic  phase ($\lambda>1$). 
The quantum phase transition is detected by a singularity of its first derivative:
\begin{equation}
\partial_{\lambda} \tau(\lambda) \stackrel{\lambda \rightarrow 1^+}{\sim} -{\rm cost}\times(\lambda-1)^{-1/4}.
\end{equation}

\section{Block entropy}
\label{sec:entropy}

In this section we compute the entanglement of a block of contiguous spins. 
As a measure of entanglement, we take the Von Neumann entropy:
\begin{equation}
S_L={\rm Tr} \left( \rho_{\{L\}} \log_2 \rho_{\{L\}} \right),
\label{blockentropy}
\end{equation}
where $\rho_{\{L\}}$ is the density matrix of a block of length $L$ defined in Eq.~(\ref{eq:block_expansion}) and does not depend on the position of the block because of the translational invariance of the system.

To calculate $S_L$ we will follow the procedure originally developed in  Ref.~\cite{kitaev}. It results 
\begin{equation}
S_L=\sum_{j=1}^L H\left(\frac{1+\nu_j}{2}\right),
\end{equation}
where 
\begin{equation}
H(x)=-x\log_2 x -(1-x)\log_2 (1-x) 
\end{equation}
is the Shannon entropy,
and $\pm \ii \nu_j$, ($j=1,\dots, L$) are the  (purely imaginary) eigenvalues of 
the Majorana correlation matrix $\Gamma_L$ in Eq.~(\ref{eq:deltagamma}).

\begin{figure}
\begin{center}
\includegraphics[width=\columnwidth]{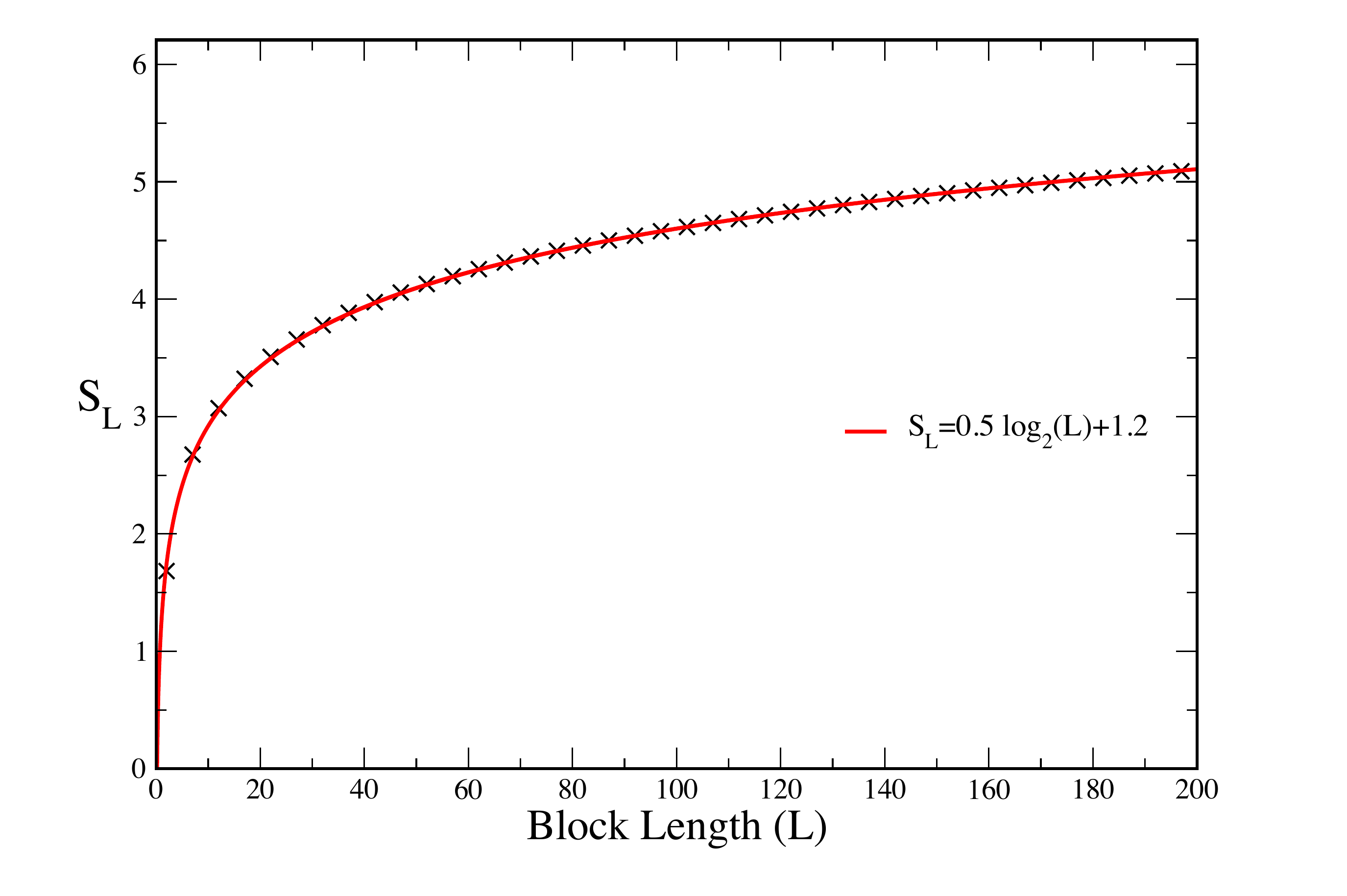}
\caption{(Color online) Block entropy $S_L$ (crosses) and its logarithmic least square fit (full line) versus the size of the block $L$. }
\label{fig:entropy}
\end{center}
\end{figure}

We have evaluated $S_L$ numerically with blocks of length ranging from 2 to 200 spins, at the critical point $\lambda=1$. The resulting behavior, displayed in Fig.~\ref{fig:entropy}, is
\begin{equation}
S_L \sim \frac{1}{2} \log_2 L+a,
\label{eq:entropyfit}
\end{equation}
where the least square fit yields for the constant multiplying the logarithmic term the value 0.506 with a standard error 0.001, while $a\simeq 1.236$ with an error 0.008. This value of $a$ is in agreement with that one would obtain by self-dual arguments
\begin{equation}
a=\frac{1+\gamma_E + (2-6\, I_3) \ln 2 - \ln 3}{2 \ln2} \simeq 1.246,
\end{equation}
where $\gamma_E \simeq 0.577$ is the Euler-Mascheroni constant
and $I_3 \simeq 0.022$ another constant \cite{zimboras}.

The logarithmic scaling of the entropy at the critical point is a result obtained also in other models  \cite{latorre}. Moreover the multiplicative constant of the logarithmic term is related to the central charge of the 1+1 dimensional conformal theory that describes the critical behavior of the chain, trough the relation  \cite{holzhey}
\begin{equation}
S_L\sim \frac{c+\bar{c}}{6} \log_2 L,
\end{equation}
where $c$ and $\bar{c}$ are the central charges of the so-called holomorphic and antiholomorphic sectors of the conformal field theory.
Also the central charge is directly related to the universality class of the system, so that, if two quantum one-dimensional models belong to the same universality class, they have the same central charge. 

The result (\ref{eq:entropyfit}) shows that the central charge of our critical chain is $c=\bar{c}=3/2$. See Table \ref{table2}. This value is different from that of the Ising chain, where $c_{\mathrm{Ising}}=1/2$, so that our model is in a different universality class than the Ising model in transverse field.

Below we exploit the periodicity of the free energy (see Sec.~\ref{sec:model}) to prove that  the CIM Hamiltonian can be indeed recast in  a sum of three decoupled Ising Hamiltonians.
From Eq.~(\ref{eq:diagonal_form}), by taking $N=3M$ and splitting the sum in three parts, we get
\begin{equation}
H(\lambda)=\sum_{s=1}^3 H_{\mathrm{Ising}}^{(s)},
\end{equation}
where
\begin{equation}
H_{\mathrm{Ising}}^{(s)} =2 \sum_{k=1}^M  \Lambda_k^{\mathrm{Ising}} \left( \gamma_k^{(s)\dagger} \gamma_{k}^{(s)}-\frac{1}{2}\right),
\label{eq:Isings}
\end{equation}
with $\gamma_k^{(s)}= \gamma_{k+(s-1)M}$ and
\begin{equation}
\Lambda_k^{\mathrm{Ising}}=\sqrt{1+\lambda^2-2 \lambda \cos\left(\frac{2 \pi k}{M} \right)}. 
\label{eq:LambdaIsing}
\end{equation}
Recalling that the central charge is an extensive quantity \cite{Di_Francesco}, we conclude that the criticality of the CIM arises as a simple sum of three Ising central 
charges.

\section{Conclusions and outlook}
\label{sec:conclusions}

We studied the statistical mechanics of the cluster-Ising Hamiltonian (\ref{eq:ham_cluster}), providing important elements for the understanding of the phase diagram of the system. The specific correlation pattern that we found indicates that most of the physics of the system is displayed beyond two-spin correlations. This is manifest in the type of entanglement encoded in the ground state that is of multipartite form.
The QPT occurring at $\lambda=1$ is exotic because the two phases enjoy distinct symmetries that cannot be continuously joined; the cluster phase does not admit any local order parameter. 

The universality class of the Cluster-Ising model (\ref{eq:ham_cluster}) is summarized in Table \ref{table2}.
Below we  compare our findings with other spin models with triplet interaction in (1+1) dimension enjoying a $Z_2\times Z_2$ symmetry and displaying a continuous phase transition \cite{griffiths,merlini}. 
The current understanding of such class of models benefits of the exact solution by Baxter and Wu \cite{baxter-wu}, corroborated by the calculation of the spontaneous magnetization by Joyce \cite{joyce}.Their critical properties  can be extracted through the analysis of  the quantum criticality of quantum spin chains with mixed two and three spin interactions in an external field \cite{penson}. 
There is general consensus that the universality class of such a class of quantum models  at zero external field is identified by the criticality of the four-state Potts model \cite{den_njis,wu_rev} at equilibrium: $\nu=2/3$, $\beta=1/12$, with some controversy on the dynamical critical exponent $z\approx 2 - 3 $ \cite{critical_z}, and central charge $c=1$.  When the external fields are switched on, the degeneracy of the ground state results to be affected in a nontrivial way. Here we mention that for a transverse fields, and for moderate longitudinal field, the resulting models are still in the universality class of the four-state Potts model; for higher longitudinal fields, criticality is shared with the three-state Potts model, yielding $c=4/5$ \cite{Blote,penson}. 
The criticality of all these models appears to be very different from the Cluser-Ising model investigated in the present  work.

Finally, it would be interesting to study this model from an experimental point of view, for instance in the context of adiabatic quantum computation. 
In this case, one could study how the encoding of the cluster phase starting from the antiferromagnetic one is affected by the critical point at different temperatures.

\acknowledgments 
We thank A. Hamma, A. Scardicchio and W. Son for discussions and Z. Zimboras for a pertinent remark.
P.F.\ and G.F.\ acknowledge support through the project IDEA of University of Bari.
R.F.\ acknowledges support from EU-IP-SOLID.

\end{document}